\documentclass[preprint,showpacs,eqsecnum,aps]{revtex4}
\usepackage{graphicx}
\usepackage{psfig}
\begin{document}
\title{New results for the two neutrino double beta decay in deformed nuclei
with angular momentum projected basis}

\author{A. A. Raduta$^{a),b)}$, C. M. Raduta$^{b)}$, A. Escuderos$^{c),d)}$}
\address{
$^{a)}$Department of Theoretical Physics and Mathematics,Bucharest University, POBox MG11,
Romania}
\address{$^{b)}$Institute of Physics and Nuclear Engineering, Bucharest, POBox MG6, Romania}
\address{$^{c)}$Instituto de Estructura de la Materia, Consejo Superior de Investigationes Cientificas,
Serrano 123, E-28006 Madrid, Spain}
\address{$^{d)}$Department of Physics and Astronomy, Rutgers University, Piscataway,
New Jersey USA 08854-8019}
\begin{abstract}
Using an angular momentum projected single particle basis, a pnQRPA approach is used
 to study the $2\nu\beta\beta$ properties of ten isotopes, exhibiting
various quadrupole deformations. The mother and  daughter nuclei exhibit
different quadrupole deformations. Since the projected basis enables a unified description of
deformed and spherical nuclei, situations where the nuclei involved in the
double beta decay process are both spherical, both deformed or one spherical and
another deformed, can be treated through a sole formalism.
Dependence of single $\beta^-$ and $\beta^+$ strength distribution on atomic mass number
and nuclear deformation is analyzed. For the double beta decay process,
the Gamow-Teller transition amplitudes and half lives are calculated.
Results are compared with the experimental data as well
as with the predictions of other theoretical approaches. The agreement
between the present results and experimental data is fairly good.
\end{abstract}

\maketitle

\section{Introduction}
\label{sec:level1}
The double beta decay of a (Z,N) nucleus, may take place through two distinct channels.
For one  channel, the final state consists of the residual nucleus
(Z+2,N-2), two electrons and two antineutrinos, while for the other one the final state
is lacking the antineutrinos. Suggestively, the two decay modes are called
two neutrinos double beta ($2\nu\beta\beta$) and neutrinoless ($0\nu\beta\beta$) double beta decay, respectively.
The second mode is especially interesting since its discovery may provide a definite answer
to the question whether neutrino is a Dirac or Majorana particle. The $2\nu\beta\beta$
process is  interesting  by its own but also due to the fact that may provide
realistic nuclear matrix elements which might be further used to quantitatively evaluate
the rate of the neutrinoless double beta decay.

For such reasons many theoreticians focused their efforts in describing
consistently the data for $2\nu\beta\beta$ decay. The contributions over
several decades have
been reviewed by many authors. Instead of enumerating the main steps achieved
toward improving the
theoretical description we advise the reader to consult a few of the review works
\cite{Suh,Ver,PriRo,HaSt,Tomo,Fass,Kla1,Rad1}.

It is interesting to note that, most of the double beta emitters are
deformed nuclei,
the proposed formalisms use a single particle spherical basis. More than 10 years ago,
one of us \cite{Rad2}
 proposed a
formalism to describe the process of two neutrinos double beta decay in a projected
spherical basis. It was for the first time that a  proton-neutron quasiparticle random
phase approximation (pnQRPA) for a
two body interaction in
the $ph$ and $pp$ channels with a deformed single particle basis was performed.
 Moreover, effects which are beyond pnQRPA, have been accounted for by means of a
 boson expansion
procedure. Few years later the influence of nuclear deformation upon the contribution
of the spin-flip
configurations to the Gamow-Teller double beta transition amplitude, was studied
\cite{Rad3}.
In the meantime several papers have been  devoted to the extension of the
pnQRPA procedure to deformed nuclei, the applications being performed for
studying the single beta decay
properties as well as the double beta decay rates.
Thus,
pnQRPA approaches using as deformed single particle basis Nilsson or deformed Woods Saxon
states have been formulated\cite{Hom,Sar,Eng,Simk}. Also, a self-consistent deformed
method was formulated where the single particle basis was obtained as eigenstates of a
deformed mean field defined through a
Hartree-Fock treatment of a density dependent two body interaction
of Skyrme type \cite{Sar}.

In a recent publication \cite{Rad04} we continued the project opened in
Ref.\cite{Rad2}, by improving the
single particle basis. Indeed, in Ref.\cite{Rad2} the single particle energies
were depending linearly on a parameter which simulates the nuclear deformation.
By contrast, in Ref. \cite{Rad04} the core  volume conservation constraint, ignored in the previous paper,
determines a nonlinear deformation dependence for single particle energies.
Of course, having  different single particle energies, the pairing
properties and the double beta matrix elements are expected to be modified.
Another issue addressed in the previous paper was whether considering different
deformations for the mother  and daughter nuclei, modifies significantly the
double beta transition amplitude (M$_{GT}$). The answer to this question is
positive since modifying the deformation for the daughter nucleus, the ground
state correlations are
modified and consequently the pnQRPA collapse point is changed. On the other
hand the overlap matrix elements of the states describing the intermediate odd-odd nucleus,
defined as excited states from the mother and daughter ground states respectively,
 are decreased. Therefore, considering different nuclear deformations for mother and
daughter nuclei quenches the Gamow-Teller (GT) double beta decay amplitude,
which results in improving the agreement with the experimental data.

The angular momentum projected spherical basis enables a unified description of
spherical and deformed nuclei. Here we use this virtue of the single particle basis
defined in Ref.\cite{Rad04} and try to depict the specific features of the transitions
between nuclei of similar or different nuclear shapes:
a) spherical $\to$ spherical ($^{48}$Ca $\to$ $^{48}$Ti), b) spherical $\to $deformed-prolate
($^{128}$Te $\to$ $^{128}$Xe, $^{130}$Te $\to$ $^{130}$Xe), c) spherical $\to $deformed-
oblate ($^{134}$Xe $\to$ $^{134}$Ba, $^{136}$Xe $\to$ $^{136}$Ba),
d) deformed $\to $spherical ($^{110}$Pd $\to$ $ ^{110}$Cd),
e) deformed-prolate$\to $deformed-prolate ($^{96}$Zr$\to $ $^{96}$Mo),
f) deformed-oblate$\to $ deformed-oblate ($^{100}$Mo $\to $ $^{100}$Ru, $^{104}$Ru
$\to $$^{104}$Pd, $^{116}$Cd $\to $ $^{116}$Sn).
 It is worth mentioning that except for $^{104}$Ru, $^{110}$Pd and $^{134}$Xe for all other
cases experimental data are available.

The results of the present paper will be described according to the
following plan. For the sake of a self-sustaining presentation,  a brief review of the projected spherical single
particle basis will be presented in Section II. Also, the basic equations necessary for
calculating the GT double beta transition amplitude are given.
In Section III, we discuss the results for ten double beta emitters:
$^{48}$Ca, $^{96}$Zr, $^{100}$Mo, $^{104}$Ru, $^{110}$Pd, $^{116}$Cd, $^{128}$Te,
 $^{130}$Te, $^{134}$Xe, $^{136}$Xe, for which the strength distribution for
 single $\beta^-$
  emission,  the M$_{GT}$ and half lives values for the double beta
 decay process, are presented. Also, the strength distribution for the
 $\beta^+$ decay of the corresponding daughter nuclei is presented as function
 of the pnQRPA energy.
  A short summary and concluding remarks are given in Section IV.

\section{pnQRPA treatment of the  Gt $\beta\beta$ transition amplitude}
\label{sec:level2}

\subsection{Projected single particle basis}
In Ref. \cite{Rad4}, one of us, (A.A.R.), introduced  an angular momentum projected single particle basis which seems to be appropriate
for the description of the single particle motion in a 
deformed mean field generated by the particle-core interaction. This single particle basis
has been used to study the collective M1 states in deformed nuclei \cite{Rad44}
as well as
the rate of double beta process \cite{Rad2,Rad3}. Recently, a new version
has been proposed where the deformation
dependence of single particle energies is nonlinear and therefore more
realistic \cite{Rad5,Rad6}.
The new single particle basis has been used to study the double beta
decay of deformed nuclei \cite{Rad04}.
In order to fix the necessary notations and moreover for the sake of a
self-contained presentation, we  describe briefly
the main ideas underlying the construction of the projected single
particle basis.

The single particle mean field is determined by a particle-core Hamiltonian: 
\begin{equation}
\tilde{H}=H_{sm}+H_{core}-
M\omega_0^2r^2\sum_{\lambda=0,2}\sum_{-\lambda\le\mu\le\lambda}
\alpha_{\lambda\mu}^*Y_{\lambda\mu}.
\label{hpc}
\end{equation}
where  $H_{sm}$ denotes the spherical shell model Hamiltonian, while $H_{core}$ is
a harmonic quadrupole boson ($b^+_\mu$) Hamiltonian, associated to a phenomenological core.
The interaction of the two subsystems is accounted for by 
the third term of the above equation, written in terms of the shape coordinates $\alpha_{00}, \alpha_{2\mu}$.
The quadrupole shape coordinates are related to the quadrupole boson
 operators by the canonical transformation:
\begin{equation}
\alpha_{2\mu}=\frac{1}{k\sqrt{2}}(b^{\dagger}_{2\mu}+(-)^{\mu}b_{2,-\mu}),
\label{alpha2}
\end{equation}
where $k$ is an arbitrary C number. The monopole shape coordinate is to be 
determined from the volume conservation condition.

Averaging $\tilde{H}$ on a given eigenstate of $H_{sm}$, denoted as usual by
$|nljm\rangle$,
one obtains a deformed quadrupole boson Hamiltonian whose eigenstate is an
axially symmetric
coherent state:
\begin{equation}
{\Psi}_g=exp[d(b_{20}^+-b_{20})]|0\rangle_b,
\label{psig}
\end{equation}
with $|0\rangle_b$ standing for the vacuum state of the boson operators and $d$ a real parameter 
which simulates the nuclear deformation.
On the other hand, averaging $\tilde{H}$ on ${\Psi}_g$ one obtains a single particle
mean field operator for the single particle motion, similar to the Nilsson
Hamiltonian.  Concluding, by averaging on a factor state of the particle core space
the rotational symmetry is broken and the mean field mentioned above may generate,
by diagonalization, a deformed basis for treating the many body interacting systems.
However, this standard procedure is tedious since the final many body states
should be projected over angular momentum.

Our procedure defines first a spherical basis for the particle-core system, by projecting out
the angular momentum from the deformed state
\begin{equation}
{\Psi}^{pc}_{nlj}=|nljm\rangle{\Psi}_g.
\label{psipc}
\end{equation}
The upper index appearing in the l.h. side of the above equation suggests that the product function is associated 
to the particle-core system.
The projected states are obtained, in the usual manner, by acting
on these deformed states with the projection operator
\begin{equation}
P_{MK}^I=\frac{2I+1}{8\pi^2}\int{D_{MK}^I}^*(\Omega)\hat{R}(\Omega)
d\Omega .
\label{pjmk}
\end{equation}
We consider the subset of projected states :
\begin{equation}
\Phi_{nlj}^{IM}(d)={\cal N}_{nlj}^IP_{MI}^I[|nljI\rangle\Psi_g]\equiv
{\cal N}_{nlj}^I\Psi_{nlj}^{IM}(d) .
\label{phiim}
\end{equation}
which are orthonormalized and form a basis for the particle-core system.
The main properties of these projected spherical states are:

a) They are orthogonal with respect to I and M quantum numbers.

b) Although the projected states are associated to the particle-core system,
they can be used as a single particle basis. Indeed, when a matrix element of 
a particle like operator is calculated, 
the integration on the core collective coordinates is performed first,  which 
results in obtaining a final 
factorized expression: one factor carries the dependence on deformation and
one is a spherical shell model matrix element.
Thus, the role of the core component is to induce a quadrupole deformation for the matrix elements of the operators acting on particle
degrees of freedom.

c) The connection between the nuclear deformation and the parameter $d$ 
entering the definition of the coherent state (\ref{psig}) can be obtained by
 requiring that the strength of the particle-core 
quadrupole-quadrupole interaction 
be identical to the Nilsson deformed term of the mean field.

To the projected spherical states, one associates the 'deformed' single particle energies
defined as  average values of the particle-core Hamiltonian
$H'=\tilde{H}-H_{core}$.
\begin{eqnarray}
\epsilon_{nlj}^I&=&\langle\Phi_{nlj}^{IM}(d)|H'|\Phi_{nlj}^{IM}(d)\rangle.
\label{epsI}
\end{eqnarray}

Since the core contribution to this average value, does not depend on the quantum numbers of
the single particle energy levels, it  produces a constant shift for all energies.
For this reason such a term is omitted in
(\ref{epsI}). However, when the ground state energy variation
against deformation is studied, this term must be included.

 In Ref.\cite{Rad04} it was shown that single particle energies, defined above,
 exhibit a  nonlinear dependence on the deformation parameter $d$. Such a dependence is
 determined by the monopole-monopole interaction term after implementing the
 volume conservation constraint.
 Moreover, the deformation dependence of the new single particle energies
 is similar to that shown by
the Nilsson model \cite{Nils}. Therefore, the average values
$\epsilon_{nlj}^I$ may be viewed as approximate expressions for the single
particle energies in deformed Nilsson orbits \cite{Nils}.
We may account for the deviations
from the exact eigenvalues by considering, 
at a later stage  when a specific
treatment of the many body system is  performed, the exact matrix elements of
the two body interaction.

Although the energy levels are similar to  those of the Nilsson model, the 
quantum numbers in the two schemes 
are different. Indeed, here we generate from each j a multiplet of $(2j+1)$
states distinguished by the quantum number $I$, which plays the role of the
Nilsson quantum number $\Omega$ and runs from 1/2 to $j$. Moreover, the energies
corresponding to the quantum numbers $K$ and $-K$ are equal to each other.
On the other hand, for a given $I$ there are $2I+1$ degenerate sub-states while the Nilsson states are only double degenerate.
As explained in Ref.\cite{Rad4}, the redundancy
problem can be solved by changing the normalization of the model functions:
\begin{equation}
\langle\Phi_{\alpha}^{I M}|\Phi_{\alpha}^{I M}\rangle=1 \Longrightarrow \sum_{M}\langle\Phi_{\alpha}^{IM}|\Phi_{\alpha}^{IM}\rangle=2.
\label{newnorm}
\end{equation}
Due to this weighting factor the particle density function is providing the
consistency result that the number of particles which can be
distributed on the (2I+1) sub-states is at most 2, which agrees with the
Nilsson model.
Here $\alpha$ stands for the set of shell model quantum numbers $nlj$.
Due to this normalization, the states $\Phi^{IM}_{\alpha}$ used to calculate the
matrix elements of a given operator should be multiplied with the weighting factor $\sqrt{2/(2I+1)}$.

       Concluding, the projected single particle basis is defined by Eq. (\ref{phiim}).
Although these states are associated to a particle-core system, they can be used as a single particle basis due to the properties 
mentioned above.
The projected states might be thought of as eigenstates of an effective rotational invariant fermionic one-body
Hamiltonian $H_{eff}$, with the corresponding  energies given by Eq.(\ref{epsI}).
\begin{equation}
H_{eff}\Phi^{IM}_{\alpha}=\epsilon^I_{\alpha}(d)\Phi^{IM}_{\alpha}.
\label{Heff}
\end{equation}
This definition should be supplemented by the request that the matrix elements of any operator between states $\Phi^{IM}_{\alpha}$ and 
$\Phi^{I^{\prime}M^{\prime}}_{\alpha^{\prime}}$ have, as we mentioned above,
a factorized form, one factor carrying the $d$ dependence, while the second one
being a spherical shell model matrix element.
Due to these features, these states can be used as single particle basis to
treat many body Hamiltonians which involve one-body operators.
This is the case of Hamiltonians with two body separable forces. As a matter of fact,
such a type of Hamiltonian is used in the present paper.

As shown in Ref. \cite{Rad04} in the
vibrational limit, $d\to 0$, the projected spherical basis goes to the spherical shell
model basis and $\epsilon_{nlj}^I$ to the eigenvalues of $H_{sm}$.

A
fundamental result obtained in Ref.\cite{Rad6} for the product of two
single particle states, which comprises a product of two core components, deserves
to be mentioned.
Therein we have proved that the
matrix elements of a two body interaction corresponding to the present scheme are
very close to the matrix elements corresponding to spherical
states projected from  a  deformed product state with one factor as  a product of
two spherical single
  particle states, and a second factor consisting of a common collective
  core wave function. The small discrepancies of the two types of matrix elements
  could be washed out by using  slightly different strengths for the two body
  interaction in the two
  methods. Due to this property the  basis (2.6) might be used for studying any two-body interaction.

\subsection{The model Hamiltonian and its pnQRPA approach} 
In the present work we aim at describing
the Gamow-Teller two neutrino double beta decay processes with the property that
mother and daughter nuclei may exhibit different shapes. Indeed, in the chosen cases
they might be both spherical, both deformed but with different deformations or one
spherical and the other one of a deformed shape. The specific feature of the formalism
used in the present work consists in treating all cases in an unified manner by using
a sole single particle basis.

The main ingredients of our formalism are as follows.
The Fermi transitions contributing about 20\% and the
``forbidden''
transitions are ignored, which is a reasonable approximation for
the two neutrino double beta decay in medium and heavy nuclei.
As usual, the $2\nu\beta\beta$ process is conceived as two successive
single
$\beta^-$ transitions. The first transition connects the ground state of
the mother nucleus to a magnetic dipole state $1^+$ of the intermediate
odd-odd nucleus which
subsequently decays to the ground state of the daughter nucleus.
The states mentioned above, involved in the $2\nu\beta\beta$
 process, are described in the framework of the pnQRPA formalism, by using
 the following many body Hamiltonian:

\begin{eqnarray}
H=&&\sum\ \frac{2}{2I+1}(\epsilon_{\tau\alpha I}-
\lambda_{\tau\alpha})c^{\dagger}_{\tau\alpha IM}c_{\tau \alpha IM}-\sum\frac{G_{\tau}}{4}
P^{\dagger}_{\tau \alpha I}P_{\tau\alpha I'}\nonumber \\
 &+& 2\chi\sum\beta^-_{\mu}(pn)\beta^+_{-\mu}(p'n')(-)^{\mu}
 -2\chi_1\sum P^-_{1\mu}(pn)P^+_{-\mu}(p'n')(-)^{\mu}.
\label{Has}
\end{eqnarray}
The operator $c^{\dagger}_{\tau\alpha IM}(c_{\tau\alpha IM})$
creates (annihilates) a particle of type $\tau$ (=p,n)
in the state $\Phi^{IM}_{\alpha}$, when acting on the vacuum
state $|0\rangle$. In order to simplify the notations, hereafter the set of
quantum numbers $\alpha(=nlj)$ will be omitted. The two body interaction
consists of three terms, the pairing, the dipole-dipole particle-hole ($ph$) and
the particle-particle ($pp$) interactions. The corresponding strengths are
denoted by $G_{\tau} (\tau=p,n), \chi, \chi_1$, respectively. All of them are separable
interactions, with the factors defined by the following expressions:

\begin{eqnarray}
P^{\dagger}_{\tau I}&=&\sum_{M}\frac{2}{2I+1}c^{\dagger}_{\tau IM}
c^{\dagger}_{\widetilde{\tau IM}},\nonumber\\
\beta^-_{\mu}(pn)&=&\sum_{M,M'}\frac{\sqrt{2}}{{\hat I}}
\langle pIM|\sigma_{\mu}|n I'M'\rangle \frac{\sqrt{2}}{{\hat {I'}}}
c^{\dagger}_{pIM}c_{nI'M'},\nonumber\\
P^-_{1\mu}(pn)& = & \sum_{M,M'} \frac{\sqrt{2}}{{\hat I}}\langle pIM|\sigma_{\mu}|nI'M'\rangle \frac{\sqrt{2}}{{\hat {I'}}}
c^{\dagger}_{pIM}c^{\dagger}_{\widetilde{nI'M'}}.
\label{Psibeta}
\end{eqnarray}
The remaining operators from Eq.(\ref{Has}) can be obtained from the above
defined operators, by hermitian conjugation.

The one body term and the pairing interaction terms are treated first through
the standard BCS formalism and consequently replaced by the quasiparticle
one body term $\sum_{\tau IM}E_{\tau}a^{\dagger}_{\tau IM}a_{\tau IM}$.
In terms of quasiparticle creation ($a^{\dagger}_{\tau IM}$) and annihilation
($a_{\tau IM}$) operators, related to the particle operators by means of the
Bogoliubov-Valatin transformation, the two body interaction terms, involved
in the model Hamiltonian, can be expressed just by replacing the
operators (\ref{Psibeta}) by their quasiparticle images which at their turn can be
expressed as linear combination of
 dipole two quasiparticle and quasiparticle density operators defined as:

\begin{eqnarray}
A^{\dagger}_{1\mu}(pn)&=&\sum_{m_p,m_n}C^{I_p\; I_n \; 1}_{m_p \; m_n \; \mu}
a^{\dagger}_{pI_pm_p}a^{\dagger}_{n I_n m_n},\nonumber\\
B^{\dagger}_{1\mu}(pn)&=&\sum_{m_p,m_n}C^{I_p\; I_n \; 1}_{m_p \;-m_n \; \mu}
a^{\dagger}_{pI_pm_p}a_{n I_n m_n}(-)^{I_n-m_n}=-[a^{\dagger}_{pI_p}a_{\widetilde{nI_n}}]_{1\mu}.
\label{A1B1}
\end{eqnarray}

The quasiparticle Hamiltonian is further treated within the
pnQRPA formalism, i.e. one
determines the operator
\begin{equation}
\Gamma^{\dagger}_{1\mu}=\sum_{k}[X(k)A^{\dagger}_{1\mu}(k)-Y(k)A_{1,-\mu}(k)(-)^{1-\mu}],
\label{Gama}
\end{equation}

which satisfies the restrictions:

\begin{equation}
[\Gamma_{1\mu},\Gamma^{\dagger}_{1\mu'}]=\delta_{\mu,\mu'},\;\; [H_{qp},\Gamma^{\dagger}_{1\mu}]=\omega\Gamma^{\dagger}_{1\mu}.
\label{boscom}
\end{equation}

These operator equations yield a set of algebraic equations for the  X (usually called forward going) and Y (named back-going) amplitudes:

\begin{eqnarray}
\left(\matrix{\cal{A}&\cal{B}\cr -\cal{B}& -\cal{A}}\right)\left(\matrix{X \cr Y}\right)=\omega\left(\matrix{X \cr Y}\right),
\end{eqnarray}
\begin{equation}
\sum_{k}[|X(k)|^2-|Y(k)|^2]=1.
\label{rpaec}
\end{equation}
The analytical expressions for the pnQRPA matrices $\cal{ A}$ and $\cal{B}$
are given in Ref.\cite{Rad04}.
Since the $pp$
interaction has an attractive character, for a critical
value of $\chi_1$ the  lowest root of the pnQRPA equations may become imaginary.
Suppose that $\chi_1$ is smaller than its critical value and therefore all RPA solutions (i.e. $\omega$) are real numbers and ordered as:
\begin{equation}
\omega_1\le\omega_2\le...\le \omega_{N_s}.
\label{omordo}
\end{equation}
Here $N_s$ stands for the total number of the proton-neutron pair states whose
angular momenta can couple to $1^+$ and moreover their quantum numbers $n,l$ are the same.
Hereafter the phonon amplitudes $X$ and $Y$ will be accompanied by a lower
index ``$i$'' suggesting that they correspond to the energy $\omega_i$.

Since our single particle basis states depend on the deformation parameter $d$,
so do the pnQRPA energies and amplitudes.
The pnQRPA ground state (the vacuum state of the pnQRPA phonon operator)
describes an even-even system which might be either the mother or the daughter nucleus.
In the two cases the gauge and nuclear deformation properties are different
which results in determining distinct pnQRPA phonon operators acting on different vacua describing the mother and daughter ground states, respectively.
Therefore, one needs an additional index distinguishing the phonon operators of the mother and daughter nuclei.
The single  phonon states are defined by the equations:
\begin{equation}
|1_{k}\mu\rangle_j=\Gamma^{\dagger}_{jk;1\mu}|0\rangle_j,\;j=i,f;\; k=1,2,...N_s.
\label{rpast}
\end{equation}
Here the indices $i$ and $f$ stand for initial (mother) and final (daughter) nuclei, respectively.
This equation defines two sets of non-orthogonal states, $\{|1_k\mu\rangle _i\}$
and $\{|1_k\mu\rangle _f\}$
 describing the
neighboring odd-odd nucleus. The states of the first set may be fed by a beta minus
decay of the ground state of the mother nucleus while the states of the second set are populated with a beta plus transition operator from the ground state of the daughter nucleus.

If the energy carried by leptons in the intermediate state is approximated by
the sum of the rest energy of the emitted electron and half the Q-value of
the double beta decay process
\begin{equation}
\Delta E=\frac{1}{2}Q_{\beta\beta}+m_ec^2,
\label{DeltaE}
\end{equation} 
the reciprocal value of the $2\nu\beta\beta$ half life can be factorized as:
\begin{equation}
(T^{2\nu\beta\beta}_{1/2})^{-1}=F|M_{GT}(0^+_i\rightarrow 0^+_f)|^2,
\label{T1/2}
\end{equation}
where F is an integral on the phase space, independent of the nuclear structure,
 while M$_{GT}$ stands for the Gamow-Teller transition amplitude and has
 the expression :

\begin{equation}
M_{GT}=\sqrt{3}\sum_{kk'}\frac{_i\langle0||\beta^+_i||1_k\rangle_i 
\mbox{}_i\langle1_k|1_{k'}\rangle_f 
\mbox{}_f\langle1_{k'}||\beta^+_f||0\rangle_f}{E_k+\Delta E+E_{1^+}}.
\label{MGT}
\end{equation}
In the above equation, the denominator consists of three terms: a) $\Delta E$,
which was already defined, b) the average value of the k-th pnQRPA energy
normalized to the particular value corresponding to k=1, i.e.

\begin{equation}
E_k=\frac{1}{2}(\omega_{i,k}+\omega_{f,k})-\frac{1}{2}(\omega_{i,1}+\omega_{f,1}),
\label{Ek}
\end{equation}
and c) the experimental energy for the lowest $1^+$ state.
The indices carried by the transition operators indicate that they act in the
space spanned by the pnQRPA states associated to the initial (i) or final
(f) nucleus. The overlap matrix elements of the single phonon states in the mother and
daughter nuclei respectively, are calculated within the pnQRPA approach and
have the expressions:
\begin{equation}
_i\langle 1_k|1_{k'}\rangle _f=\sum_{pn}\left[X_{k}(i,pn)X_{k'}(f,pn)-Y_{k}(i,pn)Y_{k}(f,pn)\right].
\label{1k1k}
\end{equation}
Throughout this paper, the Rose \cite{Rose} convention
for the Wigner Eckart theorem is used.

Before closing this section we would like to say a few words about what is
specific to our formalism.
Since our single particle states are projected spherical states,
the pnQRPA formalism is fully identical to the one, usually employed for spherical
nuclei. Since in the vibrational limit, ($d\to 0$), our basis goes to the spherical
shell model basis, one may say that the present formalism provides a unified description
of spherical and deformed nuclei.

\section{Numerical results} 
\label{sec:level3}
The formalism described in the previous section, has been applied to  the following
ten isotopes: $^{48}$Ca, $^{96}$Zr, $^{100}$Mo, $^{104}$Ru, $^{110}$Pd, $^{116}$Cd,
$^{128}$Te, $^{130}$Te, $^{134}$Xe, $^{136}$Xe.
The spherical shell model parameters for these double beta emitters and the corresponding
daughter nuclei are given by:
\begin{equation}
\hbar\omega_0=41A^{1/3},\;\;C=2\hbar\omega_0\kappa,\;\;D=\hbar\omega_0\mu,
\end{equation}
with the strength parameters $\kappa$ and $\mu$ having the same (Z,N)
dependence  as  in Ref. \cite{Ring}.

The angular momentum projected basis depends on two additional parameters. These
 are the
deformation parameter $d$ and the factor $k$ entering the canonical transformation
relating the quadrupole coordinate and boson operators.
They are fixed in the same manner as in our previous publication \cite{Rad04}.
Indeed, we require that the relative energy for the states $|1f\frac{7}{2}\frac{7}{2}\rangle$
and $|1d\frac{5}{2}\frac{1}{2}\rangle$ be equal to that of Nilsson levels with
$\Omega=\frac{7}{2}$
and $\Omega=\frac{1}{2}$ in the $N=3$ major shell.
Moreover, adding to the mean field
term defined before a $QQ$ two body interaction we require that the lowest root for the
 charge conserving QRPA equation be equal to the experimental energy of the
 lowest  $2^+$ state in the mother nucleus.
 Throughout this paper, the M-degenerate states
$\Phi^{IM}_{nlj}$ are denoted by $|n+1\;ljI\rangle$.

The BCS calculation has been performed within a  restricted single particle space.
Due to the level crossing, the restriction of the single particle space for deformed nuclei
is different from that for spherical nuclei. Indeed, in spherical nuclei Ikeda sum rule (ISR) is
satisfied if two major shells plus the spin orbit partner of the intruder state
are included in the single particle space.
Suppose that the neutron open shell has N=3 with the intruder state
$|1g9/2\rangle$,
in the
standard spherical shell model picture. In the present formalism, including the spin-orbit partner state
$|1g7/2\rangle$  means to consider  the states
$\Phi^{IM}_{0,4,\frac{7}{2}}$ with
$I=7/2,\;5/2,\;3/2,\;1/2$. However, some of these states are higher in energy than
states belonging to the $|2d5/2I\rangle$ multiplet. Due to such features
appearing both in the upper part  of the
major open shell of neutrons  and the bottom side of the proton major open shell
we truncated the space considering an inert (Z,N) core and a number of states
lying above the core states. The core and the number of outside states are
chosen such that
the non-occupation probabilities for the neglected bottom states as well as the occupation
probabilities for
the ignored
upper states are smaller than 0.01. Of course, the single particle space for
protons and neutrons are the same.
Our calculations were performed with the core and number of states given in
Table I. Once the single particle space is defined, the number of the dipole proton-neutron states
can be calculated. Furthermore, the dimensions of the pnQRPA matrices for mother ($D_1$)
and daughter ($D_2$) nuclei are readily obtained. These dimensions are also given in Table I.
It is worth mentioning that using the single particle spaces given in Table I,
Ikeda sum rule is satisfied for both the mother and daughter nuclei considered in the
present paper.

\begin{table}[h!]
\begin{tabular}{|c|cccccccccc|}
\hline
Nucleus & $^{48}$Ca & $^{96}$Zr
 &  $^{100}$Mo &
 $^{104}$Ru  &
 $^{110}$Pd & $^{116}$Cd &
 $^{128}$Te &  $^{130}$Te
     & $^{134}$Xe &  $^{136}$Xe\\
\hline
The (Z,N) core&(0,0) & (20,20) & (26,26) & (26,26) & (26,26) & (26,26) & (44,44)
& (44,44)& (44,44)& (44,44)\\
\hline
Number of states&19& 20&20& 22& 23& 27& 22&23& 21& 23\\
\hline
$D_1$&118&128 &132 & 140 &154 & 166 & 142 &
 150 & 138 & 154
\\ \hline
$D_2$&115 &128 & 132 & 140 & 154 & 166 & 128 & 132 & 120 & 140
\\ \hline
\end{tabular}
\caption{
The number of single particle proton states lying above the (Z,N) core is given. The single particle space for
neutrons is identical to that for protons. $D_1$ and $D_2$ are the dimensions of the
pnQRPA matrix for mother and daughter nuclei, respectively}.
\end{table}

Note that despite the fact that single particle energies have a deformation
dependence,
we keep calling a major shell a set of states characterized by the same quantum number
$N(=2n+l)$ plus the states from the shell $N+1$ of maximum $j$.

Single particle parameters $d$ and $k$ as well as the pairing strengths,
fixed so that the mass difference of the neighboring even-even nuclei are
reproduced, are listed in Table III.

Now, let us turn our attention to the proton-neutron dipole interactions. In Ref[11]
it was suggested a simple A-dependence for these interaction strengths:
\begin{equation}
\chi=\frac{5.2}{A^{0.7}}MeV,\;\;\chi_1=\frac{0.58}{A^{0.7}}MeV.
\label{chi}
\end{equation}
We recall that this A dependence of the proton-neutron $ph$ interaction strengths
was obtained by fitting the position of the GT resonance for $^{40}$Ca, $^{90}$Zr and
$^{208}$Pb. The $pp$ interaction strength given above has been obtained by
fitting the half lives
for $Z\le 40$ nuclei against the single $\beta^+$ decay.
A certain caution, however, is necessary when these formulae are used, since the A dependence
is conditioned by  the mass region \cite{GroKla1} as well as by the single particle
 space
\cite{Mol,Bend}. For example, in Ref.\cite{Mad} the GT resonance centroids in $^{128}$Te and $^{130}$Te, located
at 13.7 and 14.1 MeV respectively, are reproduced with the $\chi$ values equal
to 0.157 and 0.16 MeV respectively. These values for $\chi$ are different from the
predictions of Eq.(\ref{chi}) corresponding to A=128 and A=130, respectively.
Moreover, as we see from Table IV,in the current paper
 the right position of these  GT resonances are obtained  
by using
$\chi=0.268$ for both isotopes.
\begin{table}[h!]
\begin{tabular}{|c|c|c|c|c|}
\hline
Mother & Transition & Intermediate& Transition & Daughter \\
nucleus& $log\;ft$  & nucleus     & $log\;ft$  & nucleus  \\ \hline
$^{100}$Mo&$\stackrel{ \beta^+/EC}{\leftarrow}$&$^{100}$Tc&
$\stackrel{ \beta^-}{\rightarrow}$&$^{100}$Ru\\
Exp.    &   4.45$^{+0.18}_{-0.30}$$\;^{f)}$    &      &   4.66$\;^{a)}$   &          \\
Th.     &   4.61 &      &   4.66   &          \\ \hline
$^{104}$Ru&$\stackrel{ \beta^+/EC}{\leftarrow}$&$^{104}$Rh&
$\stackrel{ \beta^-}{\rightarrow}$&$^{104}$Pd\\
Exp.    &   4.32$\;^{b)}$    &      &   4.55$\;^{b)}$   &   \\
Th.     &   4.20 &      &   4.62   &          \\  \hline
$^{110}$Pd&$\stackrel{ \beta^+/EC}{\leftarrow}$&$^{110}$Ag&
$\stackrel{ \beta^-}{\rightarrow}$&$^{110}$Cd\\
Exp.    &   4.08$\;^{c)}$    &      &   4.66$\;^{c)}$   &          \\
Th.     &   3.86 &      &   4.83   &          \\  \hline
$^{116}$Cd&$\stackrel{ \beta^+/EC}{\leftarrow}$&$^{116}$In&
$\stackrel{ \beta^-}{\rightarrow}$&$^{116}$Sn\\
Exp.    &   4.39$^{+0.1}_{-0.15}$ $\;^{g)}$   &      &   4.662 $\;^{d)}$   &          \\
Th.     &   4.05 &      &   4.670  &          \\  \hline
$^{128}$Te&$\stackrel{ \beta^+/EC}{\leftarrow}$&$^{128}$I&
$\stackrel{ \beta^-}{\rightarrow}$&$^{128}$Xe\\
Exp.    &   5.049 $\;^{h)}$  &      &   6.061 $\;^{e)}$   &          \\
Th.     &   4.930 &      &   6.226  &          \\  \hline
\end{tabular}
\caption{The experimental and theoretical  $log\; ft$ values characterizing
the $\beta^+/EC$ and $\beta^-$ processes
of the intermediate nucleus ground state ($1^+$). Experimental data are from: $^{a)}$\cite{Bal},$^{b)}$\cite{Jea},
$^{c)}$\cite{Fre}, $^{d)}$\cite{Jea1}, $^{e)}$\cite{Kan}, $\;^{f)}$\cite{Garc},
 $\;^{g)}$\cite{Bhatt}, $\;^{h)}$\cite{Leder} }.
\end{table}
It is noteworthy the fact that  the daughter nuclei involved in a double beta process are stable against
$\beta^+$ transitions. Therefore $\chi_1$ is to be determined either using
information about the half life of a $\beta^+$ emitter lying close, in the nuclide chart,
to the daughter nucleus under consideration or by fitting the data for
a (p,n) reaction having the daughter as a residual nucleus.  Hereafter, the ratio $\chi/\chi_1$ is denoted, as usual, by $g_{pp}$.

The adopted procedure to fix the proton-neutron dipole interaction strengths
is as follows. Whenever, in the intermediate odd-odd nucleus, the position of
the GT resonance centroid is known, the $ph$ interaction strength is fixed so that
the above mentioned data is reproduced. As shown in Table II,
for some of the isotopes considered in the present paper
the $log\;ft$ values associated to the $\beta^+/EC$ and $\beta^-$ transitions 
of the corresponding intermediate nuclei, are experimentally known. For these particular cases,
$\chi$ and $g_{pp}$ are fixed by fitting the two mentioned experimental data.
The $log\;ft$ values were calculated by using the following expression for $ft$:
\begin{equation}
ft_{\mp}=\frac{6160}{[ {_l}\langle 1_1||\beta^{\pm}||0\rangle_l g_A]^2}.
\end{equation}
Here $|1_1M\rangle$ denotes the first dipole phonon state in the intermediate odd-odd nucleus
while $|0\rangle$ is the pnQRPA ground state. The low index "$l$" may take the value
"$i$" and "$f$" depending whether the end state of the transition is characterizing
the double beta mother or  daughter nucleus. Therefore $l=f$ is associated to single $\beta^-$
transition, while $l=i$ to the $\beta^+/EC$ process. We chose $g_A=1.0$ in order
to take account of the effect
of distant states responsible for the "missing strength" in the giant GT resonance \cite{Suh}.
For $^{48}$Ca, we considered first $\chi$ and $g_{pp}$ (second row of Table IV) as given by Eq.(\ref{chi}).
To see the effect of $g_{pp}$ on $M_{GT}$ we repeated the calculations by keeping the same
$\chi$ as before but taking $g_{pp}=0$(third row of Table IV). It seems that fixing
$\chi$ as to reproduce the GT resonance centroid and taking $g_{pp}=0$ yields a better agreement with the
experimental data. This situation is presented in the first row of Table IV.
For $^{96}$Zr, $\chi$ was fixed by fitting the
energy for the GT resonance centroid, while $g_{pp}$ was taken as required by
Eq.\ref{chi}.
For $^{130}$Te we took the same $\chi$ and $g_{pp}$ as for $^{128}$Te. It is interesting to note that for this value of $\chi$ the position of the GT resonance,
at 14.1 Mev  is nicely reproduced.
As for the last two double beta emitters included in Table IV, there are available data neither for the GT resonance nor for the $log\;ft$ values characterizing the  single $\beta^-$ and $\beta^+/EC$ transitions of the corresponding intermediate odd-odd nuclei. For these isotopes we supposed for $\chi$ and $g_{pp}$ a similar linear $1/A$ dependence as for $^{130}$Te.

The strength parameters $\chi$ and $g_{pp}$ determined in the manner described above
are collected in Table III. They are also listed for each isotope in the first row of
Table IV. These parameters yield  double beta half-lives
which are to be compared with the corresponding experimental data.
The same parameters are used to calculate the single beta strength distributions,
 shown in Figs 1-4.
 However, in order to have a fair comparison
of the present results and those of Klapdor {\it et al.}\cite{GroKla,Zha,Hir},
 in the second row of Table IV we give
the results obtained with $\chi$ and $\chi_1$  given by Eq.(\ref{chi}).

Once the parameters involved in the model Hamiltonian are fixed,
the BCS and pnQRPA equations can be solved and the results be used in
Eq.(\ref{MGT})
to calculate the $M_{GT}$ amplitude. Further,  Eq.(\ref{T1/2}) is used to calculate the
half life of the $2\nu\beta\beta$ process. The phase factor $F$ is not depending on the
nuclear state structure and therefore was taken as in Refs.\cite{Suh,Klap3}.
The values for F, used in this paper, correspond to $g_A=1.254$ (see the comments at the end of this section). Results for $M_{GT}$
and $T_{1/2}$ are given in Table IV. Therein one may find also the available experimental
data.

\begin{table}[h!]
\begin{tabular}{|c||c|c|c|c|c|c|c|}
\hline 
Nucleus & d & k & G$_{\rm p}$ [MeV] & G$_{\rm n}$ [MeV] & $\chi$ [MeV]
& g$_{\rm pp}$&$\left(\frac{1}{2}Q_{\beta\beta}+m_ec^2\right)[MeV]$ \\
\hline \hline
$^{48}$Ca & 0.3 & 10.00 & 0.65 & 0.45 & 0.180 & 0.0&2.646 \\
$^{48}$Ti &0.05 &2.00 & 0.46 & 0.36 &0.180 &0.0& \\ \hline
$^{96}$Zr & 1.5 & 10.20 & 0.26 & 0.26 & 0.5 & 0.112&2.186 \\
$^{96}$Mo & 1.2 &7.20 & 0.3 & 0.3 &0.5 &0.112& \\ \hline
$^{100}$Mo &-1.4 &10.00 & 0.28 & 0.26 &0.060 &1.600&2.026 \\
$^{100}$Ru & -0.6 & 3.6 & 0.285 & 0.220 & 0.060 & 1.600& \\ \hline
$^{104}$Ru & -1.55 & 8.80 & 0.26& 0.2 & 0.150 & 2.750&1.161 \\
$^{104}$Pd &-1.35 &6.94 & 0.26 & 0.180 &0.150 &2.750& \\ \hline
$^{110}$Pd & -1.6 & 6.00 & 0.30 & 0.32 & 0.148 & 2.450&1.516 \\
$^{110}$Cd &-0.8 &3.06 & 0.30 & 0.18 &0.148 &2.450& \\        \hline
$^{116}$Cd & -1.8 & 3.00 & 0.20 & 0.245 & 0.238 & 1.680&1.916 \\
$^{116}$Sn &-1.2 & 2.50 & 0.18 &0.275 &0.238&1.680& \\          \hline
$^{128}$Te & 0.5 & 1.62 & 0.27 & 0.220 & 0.268 & 1.250&0.946 \\
$^{128}$Xe &1.7 &6.50 & 0.23 & 0.220 &0.268 &1.250& \\       \hline
$^{130}$Te &0.493&1.88&0.24&0.210&0.268&1.300&1.776\\
$^{130}$Xe &1.4  & 5.00&0.24&0.205&0.268&1.300&    \\        \hline
$^{134}$Xe &-0.1 & 1.95 & 0.28&0.300&0.260&1.261&0.931\\
$^{134}$Ba &-0.468 &1.50 & 0.24&0.240&0.260&1.261&     \\       \hline
$^{136}$Xe &-0.1 & 1.80 &0.23&0.29&0.256&1.243&1.751\\
$^{136}$Ba &-0.698&2.16&0.19&0.20&0.256&1.243& \\ \hline
\end{tabular}
\caption{ The pairing and Gamow Teller  $ph$ interaction strengths are given
in units of MeV. The ratio of the two dipole interaction
( particle-hole and particle-particle) strengths, denoted by $g_{pp}$, is also
given.
The list of the deformation parameter $d$ and the factor $k$ of the transformation
(2.2) are  also presented.
The manner in which these parameters were fixed is explained in the text.}
\end{table}
\clearpage

\begin{table}[h!]
\begin{tabular}{|c||c|c|c|c|c|c|c|}
\hline 
$2\nu\beta\beta$ decay & $\chi$ & g$_{\rm pp}$ & $|$M$_{\rm GT}$$|$ & \multicolumn{4}{c|}{T
$_{1/2}$ [yr]} \\ \cline{5-8}
        & [MeV]       &              & [MeV$^{-1}]$             & present &exp.&
        Suhonen et al.&Klapdor et al.\\
\hline \hline
$^{48}$Ca $\to ^{48}$Ti & 0.180 & 0.0 & 0.043 & $5.23 \cdot 10^{19}$ &
 (4.2$\pm1.2)\cdot 10^{19}$ $ ^{\rm a)} $&& $3.2 \cdot 10^{19}$ $ ^{ 1)}$ \\
 & 0.346 & 0.112 & 0.032 & $9.27 \cdot 10^{19}$ & & & \\
 & 0.346 & 0.0 & 0.036      & $7.48 \cdot 10^{19}$ & & &  \\
 \hline
$^{96}$Zr $\to ^{96}$Mo & 0.500 & 0.112 & 0.113 & $1.66 \cdot 10^{19}$
&$(1.4^{+3.5}_{-0.5})\cdot 10^{19}$ $ ^{\rm a)}$ &
$0.44 \cdot 10^{20}$ $ ^{ 2)}$&$5.2\cdot 10^{17}$\\
                        &  0.213& 0.112  & 0.219 & $0.44 \cdot 10^{19}$&
                        & &\\ \hline
$^{100}$Mo $\to ^{100}$Ru & 0.060 & 1.600 & 0.305 & $4.61 \cdot 10^{18}$ &
  $(8.0\pm0.6)\cdot 10^{18}$ $ ^{a)}$&$2.9\cdot 10^{18}$ $ ^{ 3)}$&$1.8\cdot 10^{18}$  \\
&0.207&0.112&0.212&$9.55\cdot 10^{18}$& $(0.115^{+0.03}_{-0.02})\cdot 10^{20}$
                          $ ^{\rm b)}$& & \\
                          & & & & &$(0.033^{+0.02}_{-0.01})\cdot 10^{20}$
                          $ ^{\rm c,d)}$& & \\ \hline
$^{104}$Ru $\to ^{104}$Pd & 0.150 & 2.750 & 0.781 & $0.76 \cdot 10^{21}$ & &
  &$1.8\cdot 10^{21}$ \\
                         &0.201& 0.112 & 0.343 &$3.95\cdot 10^{21}$& & &
                         $3.09 \cdot 10^{22}$ $ ^{ 6)}$\\ \hline
$^{110}$Pd $\to ^{110}$Cd & 0.148 & 2.45 & 0.263 & $15.85 \cdot 10^{19}$ &
&  &$5.0\cdot 10^{19}$ \\
                          &0.194& 0.112 & 0.218 & $22.99\cdot 10^{19}$& & &
                          $1.24 \cdot 10^{21}$ $ ^{ 6)}$ \\ \hline
$^{116}$Cd $\to ^{116}$Sn & 0.238 & 1.680 & 0.116 & $3.86 \cdot 10^{19}$ &
 $(3.2\pm 0.3) \cdot 10^{19}$ $ ^{\rm a)}$ & $5.1\cdot 10^{19}$ $ ^{ 5)}$ & $8.3\cdot 10^{18}$  \\
                          &0.187 & 0.112 & 0.069 &$ 10.96\cdot 10^{19}$ &  & &$3.75\cdot 10^{19}$
                           $ ^{4)}$ \\ \hline
$^{128}$Te $\to ^{128}$Xe & 0.268 & 1.250 & 0.090 & $0.55\cdot 10^{24}$ &
 $(7.2 \pm 0.3)\cdot 10^{24}$ $ ^{\rm a)}$ & $5.6 \cdot 10^{23}$ $ ^{ 5)}$&$1.2\cdot 10^{23}$\\
                         &0.174&0.112&0.127&$0.28 \cdot 10^{24}$&
                         $(1.5\pm0.2)\cdot 10^{24}$ $ ^{\rm e)}$& &$5.7\cdot 10^{23}$ $ ^{ *)}$
                                               \\ \hline
$^{130}$Te $\to ^{130}$Xe & 0.268 & 1.300 & 0.055 & $0.261 \cdot 10^{21}$ &
 $(1.5-2.8)\cdot 10^{21}$ $ ^{\rm b)}$& $0.26 \cdot 10^{21}$ $ ^{ 5)}$&$1.9\cdot 10^{19}$\\
                         &0.172&0.112&0.091&$0.097 \cdot 10^{21}$&
                         $(2.7\pm0.1)\cdot 10^{21}$ $ ^{\rm a)}$ & &$1.2\cdot 10^{20}$ $ ^{ *)}$\\
  & & & & & $(0.75\pm 0.3)\cdot 10^{21}$ $ ^{\rm f)}$& &  \\ \hline
$^{134}$Xe $\to ^{134}$Ba & 0.260 & 1.261 & 0.039 & $3.75 \cdot 10^{24}$ &
 & & $5.1\cdot 10^{22}$ \\
                          &0.169 & 0.112 & 0.040 & $3.49\cdot 10^{24}$& & &
                          $2.5\cdot 10^{23}$ $ ^{ *)}$\\ \hline
$^{136}$Xe $\to ^{136}$Ba & 0.256 & 1.243 & 0.039 & $5.102 \cdot 10^{20}$ &
$>8.1\cdot 10^{20}$ $ ^{\rm a)}$ &$1.3\cdot 10^{20}$ $ ^{ 5)}$ & $6.0\cdot 10^{19}$ \\
                          &0.167 & 0.112 & 0.068 & $1.69\cdot 10^{20}$& & &
                          $3.3\cdot 10^{19}$ $ ^{ *)}$\\ \hline
\end{tabular}
\caption{\small{\small{ The Gamow-Teller amplitude for the $2\nu\beta\beta$ decay, in units of
MeV$^{-1}$,
and the corresponding  half life ($T_{1/2}$), in units of $yr$, are listed for
ten ground to ground
transitions. The experimental half lives for the transitions of
$^{48}$Ca($^{\rm a)}$Ref.\cite{Vog}),
$^{96}$Zr
($^{\rm a)}$ Ref. \cite{Vog}) ,
 $^{100}$Mo ( $^{\rm a)}$ Ref.\cite{Vog},
 $^{\rm b)}$ Ref.\cite{Eji} $^{\rm c)}$ Ref.\cite{Kir},
$^{\rm c)}$ Ref.\cite{Vas})),
$^{116}$Cd ($^{\rm a)}$ Ref.\cite{Vog}),
$^{128}$Te ($^{\rm a)}$ Ref.\cite{Vog}, $^{\rm e)}$ Ref.\cite{Hen}),                               )
$^{130}$Te ($^{\rm a)}$ Refs.\cite{Vog}, $^{\rm b)}$ Refs.\cite{Eji},
$^{\rm f)}$ Ref.\cite{Li}),
$^{136}$Xe ($^{\rm a)}$ Refs.\cite{Vog})
are also given. In the second last column the results reported in Refs.\cite{Suh} $ ^{2)}$,
\cite{Su2} $ ^{3)}$ and \cite{Au} $ ^{5)}$ are given.
Comparison is also possible with the theoretical results from the last column
reported in Refs.\cite{GroKla} (unmarked),\cite{Zha} ($ ^{1)}$),
\cite{Hir} ($ ^{6)}$),
and \cite{KlaGro,Klap4} ($ ^{*)}$). The parameters $\chi$ and $g_{pp}$
are also given.}}}
\end{table}
\clearpage
\noindent

Before discussing the results presented in Table IV, we would like to discuss
the strength distribution
for single $\beta^-$ and $\beta^+$ transitions of mother and daughter nuclei, respectively.
Thus, in Figs. 1-4 the strengths $B'(GT)_-$ and $B'(GT)_+$ for mother and daughter nuclei respectively
, folded with a gaussian having the width equal to 1 MeV, are plotted as function of pnQRPA energies.
These are equal to one third of the $\beta^-$ and $\beta^+$ strengths
 respectively, in the standard definition.
Thus, the difference between the total strengths $B'(GT)_-$ and $B'(GT)_+$,
characterizing  the mother nucleus, is to be compared with the sum rule (N-Z).
The results of our calculations are to be compared with the available data
for the GT giant resonance and single beta strengths given in Refs.\cite{Mad,Aki,Ander}.
At a glance one may see that while for Te and Xe isotopes most of the strength is concentrated in a narrow
resonance, for the remaining nuclei the GT resonaces have a complex structure being spread over
a large energy interval. Actually this feature is in agreement with experimental data
 showing in $^{128}$I and $^{130}$I a single peak at 13.7 and 14.1 MeV respectively
 \cite{Mad} while in $^{100}$Tc and $^{116}$In \cite{Aki} two peaks  at
 (7.8,13.2) and (8.8, 14.30) MeV. The location of the above mentioned peaks are reasonable
 well reproduced. In the case of $^{100}$Mo, the first peak is higher than that centered at 12.3 MeV.
Increasing the value of the repulsive $ph$ interaction $\chi$, the ordering of the
two peaks magnitudes is changed. 

The $\beta^-$ and $\beta^+$ strengths of
$^{48}$Ca have been studied in Refs.\cite{Ander} and \cite{Aki}.
Thus,  the GT resonance has been populated
in the reaction $^{48}$Ca(p,n)$^{48}$Sc. This resonance is spread over an energy interval
between 4.5 and 14.5 MeV. The result presented for $B'(GT)_-$ in Fig.1 upper left
panel, agrees with the
experimental data. The total $\beta^-$ strength quenched
with a factor of 0.6 \cite{Zam},
accounting for the polarization effects on the single beta transition operator,
ignored in the present paper,
is compared with the corresponding data in Table V. As shown therein, the agreement
between the calculated strength and the corresponding data is reasonably good.
The only known data for the total $\beta^+$ GT strength is for $^{48}$Ti:
\begin{equation}
\sum B(GT)_+=1.42\pm 0.2
\end{equation}
Our calculations, corresponding to the first row of Table IV, predict for
this strength the value 2.59.

Comparing the $\beta^-$ strength distribution among 2qp states with those corresponding
to pnQRPA states, one may conclude that the quasiparticle correlations accounted for by the
pnQRPA approach, favors the displacement of the strength toward higher energies. This, in fact,
is due to the repulsive character of the $ph$ interaction. As shown in Figs. 1-4,
this effect is more pronounced for $^{96}$Zr, $^{128,130}$Te and $^{134,136}$Xe.

From Table VI, it results that most of the $\beta^-$ strength
is due to the transitions
relating the proton and neutron $g$ states. By contrast in
$^{48}$Ca, $^{96}$Zr and $^{116}$Cd
the single particle decays involving $f$ states carry most of the
transition strength. Also the GT resonance peak in $^{130}$Te is determined 
mainly by the transition in the 2d state.
 While in the lighter isotopes the
transitions $\nu I\to \pi I'$ where either $I$ and $I'$ are equal to
$\frac{1}{2}$ or
$\frac{3}{2}$ prevail, in Te and Xe isotopes  the transitions $\nu \frac{7}{2} \to \pi \frac{5}{2}$
and $\nu \frac{5}{2} \to \pi \frac{7}{2}$ are dominant.

Concerning the $\beta^+$ strength distribution shown in Figs. 1-4, right panels,
the following features are to be noticed.
The magnitude of this strength is much smaller than that of $\beta^-$ shown in the left panels.
Moreover, the final  states in the $\beta^+$ process are lying in the lower
part of the spectrum,
below 7.5 MeV. This suggests that the $pp$ interaction may strongly influence
the strength distribution among these states. The sensitivity of
$\beta^+$ decay rate against the $pp$ interaction was first noticed
in Ref.\cite{Cha}. Due to this feature the $\beta\beta$ transition amplitude is also
significantly affected by increasing $g_{pp}$.
Since the $ph$ and $pp$ interactions are of different nature,
one is repulsive and the other one attractive,
one expects that the two interactions have opposite effects on the $\beta^+$
strength.
  When the $pp$ interaction is large, comparing it with the $ph$ interaction,
  the $\beta^+$ strength in the quasiparticle picture is shifted toward the
  lower states. These are the cases of $^{104}$Pd, $^{110}$Cd, $^{116}$Sn.
  When both the $ph$ and $pp$ interactions are large, the $\beta^+$ strength of
  2qp states are  very much suppressed in the pnQRPA approach.
  Such situations are met for $^{128,130}$Xe,  and $^{134,136}$Ba.

By inspecting Table VIII, we conclude that the largest $\beta^+$ strength is carried by the single
particle proton-neutron transition in the shells
 1g (for $^{100}$Ru,$^{104}$Pd,
$^{116}$Sn, $^{128}$Xe, $^{136}$Ba), 2d (for $^{96}$Mo, $^{110}$Cd), 1f (for $^{48}$Ti), 2f(for  $^{134}$Ba) and 
1h(in $^{130}$Xe).
Identifying the common 2qp configurations carrying most efficiently the
$\beta^-$ and $\beta^+$ strengths  from Table VI and VIII respectively,
one may conclude
which pnQRPA states excited from the mother and daughter  ground state
respectively,
do maximally overlap each other and therefore bring large contribution to the
double beta decay. In some of the depicted cases the state excited by the $\beta^-$ transition operator
belongs to the GT resonance. An excellent example on this line is that of
the transition $^{48}$Ca$\to $$^{48}$Ti, where the state at 10.326 excited
from the ground state of $^{48}$Ca,
and that with energy of 5.957, excited from $^{48}$Ti, have maximal overlap
due to the  2qp state $\pi (3p\frac{1}{2}\frac{1}{2}),\nu(3p\frac{3}{2}\frac{3}{2})$
which, in both $\beta^-$ and $\beta^+$ transitions, carries a large strength.

Let us now focus our attention on the GT double beta transition amplitude.
This was calculated
by means of Eq.(\ref{MGT}), where the energy shifts (\ref{DeltaE}) are those listed in Table III and
the measured values for  $1^+$ are collected in Table X.
The states, energies and overlap matrix elements involved in the $M_{GT}$ expression, were calculated
within the pnQRPA approach. The results corresponding to various sets of
proton-neutron dipole interactions, fixed in the manner explained before,
are listed in Table IV. Therein, the half lives of the
$2\nu\beta\beta$ process are also given.
The agreement with the available data is fairly good.

Comparing our results with those of Klapdor {\it et al.}\cite{GroKla,Klap4,Hir} one may say that the
 half lives predicted
by the present paper with the dipole interaction strengths given by Eq.(\ref{chi})
are, without any exception, larger by a factor ranging
from
2 ($^{100}$Mo) to 31 ($^{96}$Zr). Note that projecting out the gauge symmetry
the results for Te isotopes are close to those given here for low values of $g_{pp}$.
Comparing the results corresponding to $\chi$ and $g_{pp}$ fixed by fitting either
the GT resonance centroid energy or the $log\;ft$ value for $\beta^+/EC$
transition of the
odd-odd nucleus, and the $\log\;ft$ value of the $\beta^-$ decay ending with
the $\beta\beta$ daughter nucleus,
with those obtained with a renormalised pnQRPA equations and an adjusted Woods Saxon
single particle mean field, we note that they are close to each other.

Note that for $^{48}$Ca the  sets of ($\chi, g_{pp}$), listed in the last two rows, provide
 half lives larger than experimental
data,  which suggests that the value of $\chi$  must be smaller than required by Eq.
(\ref{chi}).  Indeed, decreasing $\chi$ to the value given in the first row of Table IV
(=1.80), the agreement between the calculated GT resonance energy
and the measured GT centroids could be improved. Moreover, the agreement with experimental data concerning
$T_{1/2}$ is also improved. By comparison one can see that the agreement quality
 obtained in the present paper is similar to that yielded by a full shell model
 calculation in Ref.\cite{Zha}.
For this set of dipole interaction strengths, the quenched total strengths of $\beta^-$
transition of $^{48}$Ca and $\beta^+$ transition of $^{48}$Ti are equal to
15.65 and 2.59
respectively. The dominant peaks in the $\beta^-$ distribution
correspond to the pnQRPA energies of 6.63 and 12.61 MeV. The carried strengths
are  1.154 and 3.344
respectively.

An interesting feature for the decay of $^{48}$Ca was pointed out by the shell model studies
\cite{Zha,Zam1,Cau}. This refers to the cumulative effect brought by the low lying
states in $^{48}$Sc, which actually yield the bulk contribution to the matrix element.
The higher $1^+$ states have a coherent destructive effect on the matrix element.
It is worth investigating these aspects within the present formalism. Indeed, in Fig.5
we plotted the transition amplitude $M_{GT}$ as a function of the upper limit of energies included in the
defining equation (2.21). In other words, for a given E the energy $E_k$ defined by Eq. (2.22)
and involved in the $M_{GT}$ expression, is restricted by $E_k\leq E$.
We note that in five energy intervals   this function is
a monotonically increasing function of E, while in the following interval the
transition
amplitude decreases when states of higher energy are added. Also one notices a
saturation effect, namely the contribution of states with energy larger than 16 MeV
is very small. One may conclude that our results concerning the behavior of the
double beta transition amplitude, are on a par with those of
 the shell model calculations. It is remarkable the fact that the maxima of $M_{GT}(E)$ and $\beta^-$
strength are reached for similar energies.

\clearpage
\begin{table}
\begin{tabular}{|c|cccccccccc|}
\hline
Nucleus &$\hskip0.1cm$ $^{48}$Ca$\hskip0.1cm$ &$\hskip0.1cm$ $^{96}$Zr
$\hskip0.1cm$ & $\hskip0.1cm$ $^{100}$Mo $\hskip0.1cm$ &
$\hskip0.1cm$ $^{104}$Ru $\hskip0.1cm$ &
$\hskip0.1cm$ $^{110}$Pd$\hskip0.1cm$ & $\hskip0.1cm$ $^{116}$Cd $\hskip0.1cm$ &
$\hskip0.1cm$ $^{128}$Te$\hskip0.1cm$ & $\hskip0.1cm$ $^{130}$Te
     $ \hskip0.1cm$ & $\hskip0.1cm$ $^{134}$Xe $\hskip0.1cm$ & $\hskip0.1cm$ $^{136}$Xe
     $\hskip0.1cm$\\
\hline
$0.6\sum{B(GT)_-}$&15.650 & 28.886 &30.040& 29.527&33.172 &38.051& 43.340 & 
47.059 & 47.028 & 50.703\\
\hline
$\sum{B(GT)^{exp}_-}$&- & - &26.690&-  & -  &32.700& 40.080 & 45.900 & -  &  - \\ \hline
\end{tabular}
\caption{ Total strengths for the Gamow-Teller $\beta^-$ transition (first row) are
compared with the available experimental data (second row). Theoretical
results are quenched
with a factor of 0.6. Data for $^{100}$Mo and $^{116}$Cd are from Ref.\cite{Aki} while those for
$^{128,130}$Te are from Ref.\cite{Mad}.}
\end{table}
\clearpage

\begin{table}
\vspace*{-2cm}
\begin{tabular}{|c|c|c|c|c|c|c|}
\hline 
Nucleus & \multicolumn{2}{c|}{1st peak} & \multicolumn{2}{c|}{2nd peak}
& \multicolumn{2}{c|}{3rd peak}\\ \cline{2-7}
 & Transition & Strength & Transition & Strength & Transition & Strength \\ \hline \hline
$^{48}$Ca & $\nu (3f \frac{7}{2} \frac{5}{2}) \to \pi (3f \frac{7}{2}\frac{7}{2})$ &
 1.155&
 $\nu(3p \frac{3}{2}\frac{3}{2}) \to \pi(3p \frac{3}{2}\frac{1}{2})$ & 0.519
 &$\nu(3p \frac{3}{2}\frac{3}{2}) \to \pi(3p \frac{1}{2}\frac{1}{2})$&0.700 \\
 & & & & & $\nu(3f \frac{5}{2}\frac{3}{2}) \to \pi(3f \frac{7}{2}\frac{1}{2})$ & 3.344 \\
\hline
$^{96}$Zr & $\nu(3f \frac{5}{2}\frac{5}{2}) \to \pi(3f \frac{5}{2}\frac{3}{2})$ & 0.602&
 $\nu (3f \frac{7}{2}\frac{7}{2}) \to \pi(3p \frac{5}{2}\frac{5}{2})$ & 1.040 &
 $\nu(3f\frac{5}{2}\frac{5}{2}\to \pi(3f\frac{7}{2}\frac{3}{2})$ &5.842 \\
& $\nu(4d\frac{5}{2}\frac{3}{2})\to \pi(4d\frac{5}{2}\frac{5}{2})$&0.467&
$\nu(3f\frac{5}{2}\frac{5}{2})\to \pi(3f\frac{7}{2}\frac{5}{2})$  & 2.473 &
$\nu(3f \frac{7}{2}\frac{5}{2}) \to \pi(3f \frac{5}{2}\frac{3}{2})$ & 2.212 \\
&&&$\nu(3p\frac{1}{2}\frac{1}{2})\to \pi(3p\frac{3}{2}\frac{3}{2})$ &0.786&&\\
\hline
$^{100}$Mo & $\nu(4g \frac{9}{2}\frac{7}{2}) \to \pi(4g \frac{9}{2}\frac{9}{2})$ & 4.950
 &$\nu(4d \frac{5}{2}\frac{5}{2}) \to \pi(4d \frac{3}{2}\frac{3}{2})$&  1.456&
 $\nu(4g\frac{9}{2}\frac{9}{2})\to \pi(4g\frac{7}{2}\frac{7}{2})$ &0.702 \\
 & $\nu(4g\frac{7}{2}\frac{7}{2})\to \pi(4g\frac{9}{2}\frac{9}{2})$ &0.428  &
& & 
 $\nu(3f\frac{7}{2}\frac{3}{2})\to \pi(3f\frac{5}{2}\frac{1}{2})$ &0.716\\
 &  $\nu(4d\frac{5}{2}\frac{3}{2})\to \pi(4d\frac{5}{2}\frac{1}{2})$ &1.005 
&&& &\\
 \hline
$^{104}$Ru & $\nu(4d \frac{3}{2}\frac{3}{2}) \to \pi(4d \frac{5}{2}\frac{1}{2})$ & 1.342 &
 $\nu(4g \frac{9}{2}\frac{3}{2}) \to \pi(4g \frac{7}{2}\frac{5}{2})$ & 0.885&
 $\nu(4g \frac{9}{2}\frac{9}{2}) \to \pi(4g \frac{7}{2}\frac{7}{2})$ &0.563 \\
&$\nu(4g \frac{9}{2}\frac{3}{2}) \to \pi(4g\frac{9}{2}\frac{1}{2})$ &2.722&
$\nu(4g \frac{9}{2}\frac{7}{2}) \to \pi (4g\frac{7}{2}\frac{7}{2})$&0.400&
 $\nu(3f \frac{5}{2}\frac{1}{2}) \to \pi(3f \frac{7}{2}\frac{3}{2})$ & 1.703\\
 &$\nu(5h \frac{11}{2}\frac{11}{2}) \to \pi(5h \frac{11}{2}\frac{11}{2})$&1.321
 &    &   & $\nu(3f \frac{7}{2}\frac{3}{2}) \to \pi(3f \frac{5}{2}\frac{1}{2})$
 &0.266\\
\hline
$^{110}$Pd & $\nu(4g \frac{7}{2}\frac{5}{2}) \to \pi(4g \frac{9}{2}\frac{5}{2})$ & 2.554 &
 $\nu(4g \frac{9}{2}\frac{5}{2}) \to \pi(4g \frac{7}{2}\frac{5}{2})$ & 1.028&
$ \nu(4g \frac{9}{2}\frac{7}{2}) \to \pi(4g \frac{7}{2}\frac{5}{2})$ & 1.570\\
 & $\nu (4g \frac{7}{2}\frac{5}{2}) \to \pi(4g \frac{9}{2}\frac{3}{2})$ & 1.080 &
 $\nu(4d \frac{3}{2}\frac{1}{2}) \to \pi(4d \frac{3}{2}\frac{1}{2})$ & 1.171&
$ \nu(3f \frac{5}{2}\frac{3}{2}) \to \pi(3f \frac{7}{2}\frac{1}{2})$ &0.294 \\
& $\nu (4g \frac{9}{2}\frac{5}{2}) \to \pi(4g \frac{9}{2}\frac{7}{2})$&2.705&
&   &   & \\
\hline
$^{116}$Cd
 & $\nu (4g \frac{7}{2}\frac{5}{2}) \to \pi (4g \frac{7}{2}\frac{7}{2})$ & 1.668 &
 $\nu (4g \frac{9}{2}\frac{5}{2}) \to \pi (4g \frac{7}{2}\frac{7}{2})$ & 0.769&
$\nu (3f \frac{7}{2}\frac{3}{2}) \to \pi (3f \frac{5}{2}\frac{1}{2})$ &1.121  \\
& $\nu (4d \frac{5}{2}\frac{1}{2}) \to \pi (4d \frac{3}{2}\frac{3}{2})$ &0.462 &
$\nu (4g \frac{7}{2}\frac{7}{2}) \to \pi (4g \frac{9}{2}\frac{5}{2})$ & 0.927&
$\nu (3f \frac{5}{2}\frac{3}{2}) \to \pi (3f \frac{7}{2}\frac{1}{2})$&2.647
 \\
 &  &  &$\nu (4g \frac{7}{2}\frac{3}{2}) \to \pi (4g \frac{9}{2}\frac{1}{2})$&
 1.247&$\nu (3f \frac{7}{2}\frac{3}{2}) \to \pi (3f \frac{5}{2}\frac{3}{2})$&1.971
\\ \hline
$^{128}$Te
 & $\nu(4g \frac{7}{2}\frac{1}{2}) \to \pi(4g \frac{7}{2}\frac{3}{2})$ & 0.446 &
 $\nu(4g \frac{9}{2}\frac{5}{2}) \to \pi(4g \frac{7}{2}\frac{3}{2})$ & 0.467 &
 $\nu(4g \frac{9}{2}\frac{5}{2}) \to \pi(4g \frac{7}{2}\frac{7}{2})$ & 12.483 \\
 &$\nu(4g \frac{7}{2}\frac{3}{2}) \to \pi(4g \frac{9}{2}\frac{5}{2})$ &0.380&
 $\nu(4g \frac{9}{2}\frac{3}{2}) \to \pi(4g \frac{7}{2}\frac{1}{2})$ & 0.198&
$\nu(4d \frac{5}{2}\frac{3}{2}) \to \pi(4d \frac{3}{2}\frac{3}{2})$&7.316\\
\hline
$^{130}$Te&
$\nu(4g \frac{7}{2}\frac{5}{2}) \to \pi(4g \frac{7}{2}\frac{7}{2})$ & 0.462
&$\nu(5h \frac{11}{2}\frac{1}{2}) \to \pi(5h \frac{9}{2}\frac{3}{2})$ &0.430 &
 $\nu(4g \frac{9}{2}\frac{5}{2}) \to \pi(4g \frac{7}{2}\frac{7}{2})$ & 2.543 \\
&$\nu(4d \frac{3}{2}\frac{1}{2}) \to \pi(4d \frac{5}{2}\frac{3}{2})$ &0.342&
 $\nu(4g \frac{9}{2}\frac{3}{2}) \to \pi(4g \frac{7}{2}\frac{1}{2})$ &0.291&
$\nu(4d \frac{5}{2}\frac{3}{2}) \to \pi(4d \frac{3}{2}\frac{3}{2})$ &19.205\\
\hline
$^{134}$Xe&
$\nu(5h \frac{11}{2}\frac{9}{2}) \to \pi(5h \frac{11}{2}\frac{11}{2})$ & 0.713
&$\nu(4g \frac{7}{2}\frac{5}{2}) \to \pi(4g \frac{9}{2}\frac{5}{2})$ & 0.659 &
 $\nu(4d \frac{5}{2}\frac{1}{2}) \to \pi(4d \frac{3}{2}\frac{1}{2})$ & 0.784 \\
 & & &$\nu(4d \frac{3}{2}\frac{1}{2}) \to \pi(4d \frac{3}{2}\frac{3}{2})$ &0.625&
$\nu(4g \frac{9}{2}\frac{7}{2}) \to \pi(4g \frac{9}{2}\frac{5}{2})$&21.050\\
\hline
$^{136}$Xe&
$\nu(4g \frac{7}{2}\frac{3}{2}) \to \pi(4g \frac{9}{2}\frac{5}{2})$ & 0.710
&$\nu(4d \frac{3}{2}\frac{3}{2}) \to \pi(4g \frac{3}{2}\frac{1}{2})$ & 0.392 &
 $\nu(4g \frac{9}{2}\frac{7}{2}) \to \pi(4g \frac{9}{2}\frac{5}{2})$ & 23.661 \\
&$\nu(4d\frac{3}{2}\frac{3}{2})\to \pi(4g\frac{5}{2}\frac{5}{2})$ &0.321 &
$\nu(4d \frac{5}{2}\frac{1}{2}) \to \pi(4d \frac{3}{2}\frac{1}{2})$ &0.709&&
\\ \hline
\end{tabular}
\caption{ \small{\small{The strengths carried by the pnQRPA states
contributing to
the
first, second and third (if any) peaks from the upper panels of Figs.1-4, are listed.
On the left side of these numbers, the 2qp configurations closest
in energy to the corresponding pnQRPA states, are given. This is the
dominant configuration of the chosen $pn$ phonon state. The states $\Phi^{IM}_{nlj}$
(see Eq. (2.6) are specified by the quantum numbers (NljI)  where N=2n+l. Also,
the orbital angular momentum values 0,1,2,..are mentioned by
the letters s,p,d,.., respectively. }}}
\end{table}
\clearpage

\begin{table}[!]
\begin{tabular}{|c||c|c|c|c|c|c|}
\hline 
Nucleus & \multicolumn{2}{c|}{1st peak} & \multicolumn{2}{c|}{2nd peak}
& \multicolumn{2}{c|}{3rd peak} \\ \cline{2-7}
 & pnQRPA energy & Strength & pnQRPA energy & Strength& pnQRPA energy& Strength \\ \hline \hline
$^{48}$Ca & 6.633 & 1.155 & 7.959 & 0.519&10.326&0.700 \\
 & & &  && 12.611&3.344 \\
\hline
$^{96}$Zr & 7.077 & 0.602 & 10.316 & 1.040&12.416&5.842 \\
 &7.367 &0.467 &11.633 &2.473 &13.125&2.212 \\
 &  &   & 11.901 & 0.786 & & \\
\hline
$^{100}$Mo & 5.625 & 0.428 & 9.282 & 1.456  & 11.678& 0.702 \\
           & 5.790 & 4.950      &       &        &12.104  & 0.716  \\
           & 6.452 &1.005       &       &        &         &        \\
 \hline
$^{104}$Ru & 5.181 &1.342  & 9.028 & 0.885 & 11.296  & 0.563 \\
           &5.444  &2.722  & 9.224 & 0.400 &11.670 &1.703 \\
           &6.327  &1.321  &       &       &11.900 &0.266       \\
\hline
$^{110}$Pd & 3.470 & 2.554 & 10.319 & 1.028 & 12.641 & 1.570 \\
           & 5.053 & 1.080 & 11.067 & 1.171 & 12.792 & 0.294 \\
           & 6.339 &2.705  &        &       &        & \\
\hline
$^{116}$Cd & 2.359 & 1.668 & 6.042 & 0.769&13.236&1.121 \\
           & 3.221 & 0.462  &6.378 & 0.927 &13.346 &2.647  \\
           &       &       & 7.083 & 1.247 &13.407 &1.971 \\
\hline
$^{128}$Te & 5.339 & 0.445 & 10.173  & 0.467 & 13.713 & 12.483 \\
           & 6.880 & 0.380 &10.401  & 0.198 &  14.048 &  7.316 \\
 \hline
$^{130}$Te & 6.553 & 0.462 & 10.129 & 0.430 & 13.652 & 2.543 \\
           & 7.965 & 0.342 & 10.351 & 0.291 &14.107 &  19.205 \\
\hline
$^{134}$Xe & 4.188 & 0.713 & 7.403 & 0.659 & 12.217& 0.784 \\
           &       &       & 7.761 & 0.625 & 14.865  &21.050 \\
 \hline
$^{136}$Xe & 7.545 & 0.710 & 11.437 & 0.392 & 15.359 & 23.661 \\
           &7.902 &0.321  & 12.424 &  0.709 &        &         \\
\hline
\end{tabular}
\caption{ The energies of the pnQRPA states which give the largest strength
contributions to  the peaks in Figs. 1-4, left panels.
The carried strengths are also given.}
\end{table}

\clearpage

\begin{table}[!]
\begin{tabular}{|c||c|c|c|c|c|c|}
\hline 
Nucleus & \multicolumn{2}{c|}{1st peak} & \multicolumn{2}{c|}{2nd peak} & 
\multicolumn{2}{c|}{3rd peak} \\ \cline{2-7}
 & Transition & Strength & Transition & Strength & Transition & Strength \\ \hline \hline
$^{48}$Ti
 & $\pi(3p \frac{1}{2}\frac{1}{2}) \to \nu(3p \frac{3}{2}\frac{3}{2})$ & 0.153 &
 $\pi(3f \frac{7}{2}\frac{7}{2}) \to \nu(3f \frac{5}{2}\frac{5}{2})$ & 0.307 &
$\pi(3f\frac{7}{2}\frac{3}{2})\to \nu (3f \frac{5}{2}\frac{1}{2})$&0.562 \\
\hline
$^{96}$Mo
 & $\pi(4d \frac{5}{2}\frac{1}{2}) \to \nu(4d \frac{3}{2}\frac{1}{2})$ & 0.146 &
 $\pi(4d \frac{5}{2}\frac{3}{2}) \to \nu(4d \frac{5}{2}\frac{5}{2})$ & 0.025 & &\\
\hline
$^{100}$Ru
 & $\pi(4d\frac{5}{2}\frac{5}{2})\to \nu (4g \frac{3}{2}\frac{3}{2})$ &0.405  &
 $\pi(4g \frac{9}{2}\frac{9}{2}) \to \nu(4g \frac{7}{2}\frac{7}{2})$ & 0.509  &
   & \\
 &   &    &  $\pi(4g\frac{9}{2}\frac{7}{2})\to \nu (4g \frac{7}{2}\frac{5}{2})$ &0.362& & \\
\hline
$^{104}$Pd
 & $\pi(4g \frac{9}{2}\frac{5}{2}) \to \nu(4g \frac{7}{2}\frac{7}{2})$ & 0.441 &
$\pi(4g \frac{9}{2}\frac{1}{2})\to \nu (4g\frac{9}{2}\frac{3}{2})$ &0.247 &
 $\pi(4g \frac{7}{2}\frac{5}{2})\to \nu (4g\frac{9}{2}\frac{7}{2})$&0.023  \\
 &&&$\pi(4d \frac{5}{2}\frac{1}{2})\to \nu (4d\frac{5}{2}\frac{3}{2})$& 0.110
 & & \\  \hline
$^{110}$Cd
 & $\pi(4d \frac{3}{2}\frac{3}{2}) \to \nu(4d \frac{5}{2}\frac{1}{2})$ & 0.273 &
 $\pi(4g \frac{9}{2}\frac{7}{2}) \to \nu(4g \frac{7}{2}\frac{5}{2})$ & 0.110 &
 $\pi(4g \frac{7}{2}\frac{5}{2}) \to \nu(4g \frac{9}{2}\frac{3}{2})$&0.027   \\
 & & & $\pi(4d \frac{5}{2}\frac{3}{2}) \to \nu(4d \frac{3}{2}\frac{3}{2})$& 0.325&
 &  \\ \hline
$^{116}$Sn
 & $\pi(4g \frac{7}{2}\frac{7}{2}) \to \nu(4g \frac{7}{2}\frac{5}{2})$ & 0.391 &
 $\pi(5f\frac{7}{2}\frac{7}{2}) \to \nu (5f \frac{5}{2}\frac{5}{2})$ &0.117&
 $\pi (4d \frac{5}{2}\frac{1}{2})\to \nu (4d \frac{5}{2}\frac{3}{2})$&0.019  \\
&&& $\pi (4g \frac{7}{2}\frac{7}{2})\to \nu (4g \frac{9}{2}\frac{5}{2})$&0.927
&&\\
\hline
$^{128}$Xe
 & $\pi(4d \frac{5}{2}\frac{1}{2}) \to \nu(4d \frac{3}{2}\frac{3}{2})$ & 0.011 &
 $\pi(4g \frac{9}{2}\frac{5}{2}) \to \nu(4g \frac{7}{2}\frac{5}{2})$ & 0.020 & &\\
& $\pi(5h \frac{11}{2}\frac{1}{2}) \to \nu(5h \frac{11}{2}\frac{3}{2})$ & 0.011&
$\pi(4g \frac{9}{2}\frac{7}{2}) \to \nu(4g \frac{7}{2}\frac{7}{2})$&0.041& &   \\
\hline
$^{130}$Xe
 & $\pi(4d \frac{5}{2}\frac{3}{2}) \to \nu(4d \frac{3}{2}\frac{3}{2})$ & 0.020 &
  $\pi(4g \frac{9}{2}\frac{7}{2}) \to \nu(4g \frac{7}{2}\frac{7}{2})$ &0.026 &&\\
  &$\pi(5h \frac{11}{2}\frac{1}{2}) \to \nu(5h \frac{11}{2}\frac{3}{2})$ &0.028 &   &  &   & \\
   \hline
$^{134}$Ba
 & $\pi(5f \frac{7}{2}\frac{7}{2}) \to \nu(5f \frac{5}{2}\frac{5}{2})$ & 0.106 &
  && &\\
& $\pi(4g \frac{9}{2}\frac{3}{2}) \to \nu(4g \frac{9}{2}\frac{1}{2})$ & 0.086&&&&\\
 \hline
$^{136}$Ba
 & $\pi(4g \frac{9}{2}\frac{5}{2}) \to \nu(4g \frac{7}{2}\frac{3}{2})$ & 0.120 &
  &  & &  \\
 & $\pi(4g \frac{9}{2}\frac{3}{2}) \to \nu(4g \frac{7}{2}\frac{1}{2})$  &0.124 &&&&\\
\hline
\end{tabular}
\caption{The same as in Table VI but for the right panels of Figs. 1-4.}
\end{table}
\clearpage

\begin{table}[h!]
\begin{tabular}{|c||c|c|c|c|c|c|}
\hline 
Nucleus & \multicolumn{2}{c|}{1st peak} & \multicolumn{2}{c|}{2nd peak} & 
\multicolumn{2}{c|}{3rd peak} \\ \cline{2-7}
 & pnQRPA energy & Strength & pnQRPA energy & Strength & pnQRPA energy & Strength \\ \hline \hline
$^{48}$Ti
 & 5.957 & 0.153 &
 6.940 & 0.307 &
 7.565 & 0.562 \\
\hline
$^{96}$Mo
 & 3.361 & 0.146 &
 4.224 & 0.025 & & \\
\hline
$^{100}$Ru
 &2.099 & 0.405 &3.617&0.509& &\\
 &  & &4.437&0.362& & \\
\hline
$^{104}$Pd
 & 0.863& 0.441 & 4.265
& 0.247 &10.725 &0.023  \\
&  &  &5.300& 0.110& &  \\
\hline
$^{110}$Cd
 & 1.817& 0.273 & 4.405 & 0.110 &6.873 &0.027 \\
 &      &       & 4.531 & 0.325 &      &       \\
\hline
$^{116}$Sn
 & 1.727 & 0.391 &5.718 & 0.117 & 8.461 &  0.019\\
 &       &       & 6.378& 0.927 &       &        \\
\hline
$^{128}$Xe
 & 2.554& 0.011& 5.025 & 0.041&  & \\
 & 3.246& 0.011& 5.479 & 0.020 &  & \\
\hline
$^{130}$Xe
 & 3.356 & 0.020 &4.674 &0.026 & &  \\
 & 3.446 & 0.028 &      &      & &   \\
\hline
$^{134}$Ba
 & 3.756& 0.106 &     &     &     &  \\
 &3.803 & 0.086 &     &     &     &   \\
\hline
$^{136}$Ba
 & 3.242& 0.120 &       &       & & \\
 & 3.534 &0.124       &       &       & & \\
\hline
\end{tabular}
\caption{The same as in Table VII, but for the right panels of Figs. 1-4.}
\end{table}
\clearpage

\begin{table}[h!]
\begin{tabular}{|c|cccccccccc|}
\hline
Nucleus &$\hskip0.2cm$ $^{48}$Sc$\hskip0.2cm$ &$\hskip0.2cm$ $^{96}$Nb
$\hskip0.2cm$ & $\hskip0.2cm$ $^{100}$Tc $\hskip0.2cm$ &
$\hskip0.2cm$ $^{104}$Rh $\hskip0.2cm$ &
$\hskip0.2cm$ $^{110}$Ag$\hskip0.2cm$ & $\hskip0.2cm$ $^{116}$In $\hskip0.2cm$ &
$\hskip0.2cm$ $^{128}$I$\hskip0.2cm$ & $\hskip0.2cm$ $^{130}$I
     $ \hskip0.2cm$ & $\hskip0.2cm$ $^{134}$Cs $\hskip0.2cm$ & $\hskip0.2cm$ $^{136}$Cs
     $\hskip0.2cm$\\
\hline
E$_{1^+}$[keV]&338 & 1116 & 0 & 0 & 0 & 0 & 58 & 85 & 177 & 177\\
\hline
\end{tabular}
\caption{
The experimental energies for the first 1$^+$ states in the intermediate odd-odd nuclei
are given in units of keV. Data are taken from Refs.\cite{Bur,Pek,Bal,Jea,Fre,Jea1,Kan,Bal1,Serg,Son}.
The states in
$^{48}$Sc and $^{96}$Nb, at 338 and 1116 keV respectively, are not assigned
with angular momentum and parity.
Here we, ad hoc, suppose that they
have the angular momentum equal to one and a positive  parity. For $^{136}$Cs, there is no
available data for energy levels. For this case we adopted the same excitation
energy for the state $1^+$ as in $^{134}$Cs. Also for $^{128,130}$I the energies for $1^+$
are the same as in Ref.\cite{Mad}}.
\end{table}

\begin{table}[!]
\begin{tabular}{|c||c|c|c|c|}
\hline 
Nucleus & \multicolumn{2}{c|}{SSD} & \multicolumn{2}{c|}{present}
 \\ \cline{2-5}
          &$\hskip0.5cm$ $M_{GT}$$\hskip0.5cm$  &$\hskip0.5cm$ $t_{1/2}$$\hskip0.5cm$ &$\hskip0.5cm$ $M_{GT}$$\hskip0.5cm$  &$\hskip0.5cm$ $t_{1/2}$$\hskip0.5cm$ \\
          &           &           &           &            \\
\hline
$^{100}$Mo&0.211      &5.860$ \cdot 10^{~19}$&0.305&2.82$ \cdot 10^{~19}$\\
          &           &                    &     &   \\
\hline
$^{104}$Ru&0.616&7.493$ \cdot 10^{~21}$ & 0.781 & 4.655$ \cdot 10^{~21}$\\
          &     &                   &       &        \\
\hline
$^{110}$Pd&0.551&2.208$ \cdot 10^{~20}$ & 0.263 & 9.694$ \cdot 10^{~20}$\\
          &     &                      &       &                      \\
\hline
$^{116}$Cd&0.421&1.780$ \cdot 10^{~19}$ & 0.116 & 23.63$ \cdot 10^{~19}$\\
          &      &                     &       &                       \\
\hline
$^{128}$Te&0.032&26.950$ \cdot 10^{~24}$ & 0.090 & 3.38$ \cdot 10^{~24}$\\
          &     &                      &       &                     \\
\hline
\end{tabular}
\caption{The $M_{GT}$  and $t_{1/2}$ values obtained with the single state dominance ($SSD$) hypothesis. For an easy comparison we give also the values obtained within the present formalism. By contrast to the $t_{1/2}$ values given in Table IV, here the half lives correspond to $g_A=1.$}
\end{table}
\clearpage

Before closing this Section we would like to say a few  more words about the adopted procedure for fixing the dipole proton-neutron strength parameters.  For the sake of a unitary treatment, the half lives of all double beta decaying nuclei were calculated by taking for $g_A$ the value 1.254. However, in five of the situations considered the single beta properties for the intermediate odd-odd nuclei are determined by supposing an effective value ($g_A$=1.) for the axial-vector coupling strength, which might simulate the contribution of the higher energy states. Thus, although the nuclear matrix elements as well as the proton-neutron interaction strengths are similar for double and single transitions, we considered that the two sets of properties mentioned above are influenced by different parts of the proton-neutron QRPA excitation spectrum.
In this context it is worth mentioning that some time ago Abad ${\it et\; al.}$ \cite{Aba} advanced the single state dominance hypothesis (SSDH) which asserts that for double beta decay processes where the intermediate odd-odd nucleus has the state $1^+$ as ground state, most of the contribution to the double beta matrix element comes from the the first intermediate
dipole state. If that hypothesis works, then the double beta process is dominated by two virtual and succesive single  $\beta^-$ transitions, one  from the ground state of the mother nucleus to the ground state of the intermediate odd-odd nucleus, while the other one from there to the ground state of the daughter nucleus. In the meantime the  SSDH has been considered by many authors 
\cite{Garc,Eji1,Civi1,Civi2,Sim1,Eji2,Civi3,Sim2}. As shown in Table X, four odd-odd isotopes have indeed the first $1^+$ as ground state. Moreover, there are also available  data for $\beta^-$ and $\beta^+/EC$ transitions  of $^{130}$I.
Therefore, for all five isotopes from Table II we checked the SSDH validity by keeping in Eq. (2.21) only the first term from the sum and considering the states overlap equal to unity. Also the $t_{1/2}$ values are calculated by considering 
$g_{A}=1.$ The results are compared with those obtained by summing up over all pnQRPA states, otherwise keeping  
$g_{A}=1$ in order to have a fair comparison. Results are listed in Table XI. 
From there one may conclude that indeed our calculations confirm the SSDH for $^{100}$Mo, $^{104}$Ru and $^{110}$Pd, but not for the remaining two double beta nuclei,
$^{116}$Cd and $^{128}$Te. Most likely for these cases the summation in the expression of $M_{GT}$ should be extended from one to few states.     

It is worthwhile to address the question how stable are these results against changing the dimension of the single particle basis. We checked this feature with a positive result. To be more concrete let us describe the modifications obtained for $^{110}$Pd. For this isotope we increased the dimension $D_1$ from 23 to 27, otherwise keeping the same parameters for  single particle states as before. We changed the pairing strengths in order to preserve the pairing properties, i.e. to have the gap parameters unchanged. The new $(G_p,G_n)$ for mother and 
daughter are (0.281,0.271)MeV and
(0.2795, 0.1665)MeV, respectively. The pnQRPA matrix has the dimension 
$D_2=186.$ The ISR value devates from $N-Z$ by 3\%.
 The proton-neutron interaction strengths have been changed to ($\chi,g_{pp})=(0.13735, 2.4)$MeV in order to  keep the $log\;ft$ values for the single beta transitions of $^{110}$Ag close to the experimental data. The results for these observables are 4.84 for $\beta^-$ and 3.70 for $\beta^+/EC$. The double beta transition amplitude and the $T_{1/2}$ obtained under the new circumstances, are 0.2626 and 15.881$\cdot 10^{19}yr$. One notes that these values are very close to those listed in Table IV.
One may conclude that the results are stable against enlarging the single particle space and  moreover our choice for $D_1$ is motivated by the fact that ISR is satisfied.

\section{Conclusions}
\label{sec:level4}
In the previous sections we completed the project started in Ref.\cite{Rad04}
by studying the $2\nu\beta\beta$ decay of another ten even-even nuclei exhibiting
various shapes. In the chosen cases the mother and daughter nuclei have the following shapes:
a) both are spherical, b) both deformed-prolate, c) both deformed-oblate,
d) one spherical and another deformed-prolate, e) both are near spherical but
prolate,
f) both are near spherical but oblate.
The deformations obtained for the ten isotopes are similar to those of
Ref.\cite{GroKla3,Audi}.
In some cases these are different from nuclear deformations reported in
Ref.\cite{La}. Indeed, for example in the present paper as well as in Ref.\cite{GroKla3},
the quadrupole deformation for $^{100}$Mo is negative while in Ref.\cite{La} this is positive.
Moreover, as shown in Ref.\cite{La} a negative deformation is reached in $^{106}$Mo.
The reason for this discrepancy
might be the fact that there the stationary points of the energy function are
obtained in the space of quadrupole and hexadecapole deformations while here only
the quadrupole variable is considered. Moreover, here an angular momentum
projected  single particle basis is used.

It is manifest the fact that an oblate to oblate single beta transition
is involving single particle configurations which are different from those appearing in
a prolate to prolate transition. Indeed, suppose that a certain number of
nucleons are distributed
alternatively in a prolate and an oblate single particle levels and that in
the first case the Fermi
level for neutrons is characterized by a small quantum number $I$. In this case
 the $\beta^-$ strength for the prolate to prolate transition is carried by
 single particle dipole transitions between states of low $I$ as well as of
 large $I$ but originating from the upper shell. By contrary, in the
 oblate to oblate transitions, the privileged transitions are those relating neutron and proton single
 particle states with large $I$ and those of small $I$ from the upper shell.
 Such cases can be easily identified in Tables VI and VIII. Due to the feature
 mentioned above
 the strength fragmentation is expected to be more pronounced in the oblate to oblate
 transitions. Actually such a situation is met for  $^{100}$Mo, $^{104}$Ru, $^{110}$Pd
 and $^{116}$Cd.

 The structure of the peaks seen in Figs 1,2 in the $\beta^-$ strength distribution
 of the nuclei mentioned before is as follows. In $^{100}$Mo the peaks are determined by
 transitions inside the shells  1g (1st peak) 2d (2nd peak) and 1g and 1f (3rd peak).
 For $^{104}$Ru the first two peaks are determined mainly by the transitions from the
 shell 1g while the third one by the transition from the shell 1f.
 In $^{110}$Pd the following shells are involved in the transitions contributing most
 to the three peaks: 1g (1st peak),  1g, 2d (2nd peak) and 1g, 1f (3rd peak).
 For $^{116}$Cd only one shell contributes most to any of the three dominant peaks:
 1g (1st and 2nd  peaks), 1f (3rd peak).

Note that for Te and Xe isotopes, the $\beta^-$ strength is mainly concentrated
in one pnQRPA state. These nuclei are almost spherical
(Te isotopes are soft prolate while Xe isotopes are soft oblate). Moreover,
 in the daughter nuclei the nuclear deformation has the same sign as in the corresponding mother nuclei.
In $^{128}$Te and $^{130}$Te  the dominant single  particle state $np$ transitions are
$\nu(4g\frac{9}{2}\frac{5}{2})\to \pi(4g\frac{7}{2}\frac{7}{2})$ and $\nu(4d\frac{5}{2}\frac{3}{2}\to 
\pi(4d\frac{3}{2}\frac{3}{2})$, respectively. In Xe isotopes
the $np$ single particle transition
$\nu(4g\frac{9}{2}\frac{7}{2})\to \pi(4g\frac{9}{2}\frac{5}{2})$ , prevails.

The transition amplitudes $M_{GT}$ and half lives for the $2\nu\beta\beta$
process were calculated within the pnQRPA approach by using a projected spherical basis.
The agreement with the available data is quite good. The adopted fitting procedure
for the $pn$ dipole interaction strengths yield large values for $g_{pp}$, in several cases.
These values are not far from  the critical value, where the pnQRPA breaks down.
It is an open question whether for these transitions
a good agreement with the data would be possible by keeping a small $g_{pp}$, but
accounting for higher pnQRPA effects.
Inspecting the Table IV, one can judge not only on the agreement of the present
results with the
experimental data but also on the comparison between predictions of different
theoretical approaches. Indeed, the agreement with experimental data is reasonably good.
Although they are based on different formalisms as well as different single particle states
the present results and those of Suhonen {\it et al.}\cite{Suh,Su2,Au} are not far from each other.
Comparing the results of the present paper, obtained with $\chi$ and $g_{pp}$ given
by Eq.(\ref{chi}) and the corresponding predictions from Refs.\cite{GroKla,Zha,Hir},
one notices that they are quite different.

It is worth mentioning that for $^{104}$Ru and $^{110}$Pd the $pn$ dipole interactions
are fully determined by fitting the data concerning the $\beta^+/EC$ and $\beta^-$
decay $log\;ft$ values of the ground state ($1^+$) of the intermediate nuclei
$^{104}$Rh and $^{110}$Ag, respectively. The predictions for the double beta emitter half
lives are $0.76\cdot 10^{21}yr$ and $15.85\cdot 10^{19}yr$. They are 40 and 8 times smaller than the corresponding
predictions of Ref.\cite{Hir}.
Our prediction for the half-life of $^{134}$Xe, against double beta decay, is
$3.75\cdot 10^{24}yr$ which exceeds by a factor 15 the corresponding
finding of Ref.\cite{GroKla,Klap4}.

The single state dominance is confirmed, by our formalism, to be valid for $^{100}$Mo, $^{104}$Ru and $^{110}$Pd. 

Finally we may conclude that the projected spherical single particle basis provides a suitable framework for a unified description of the double beta properties of spherical and deformed nuclei. 
The results presented in the previous publication \cite{Rad04} and here
constitute a good starting point for studying the higher pnQRPA contributions to
the $2\nu\beta\beta$ process as well as the transitions populating the daughter nuclei in
an excited state.

\begin{figure}[h]
\centerline{\psfig{figure=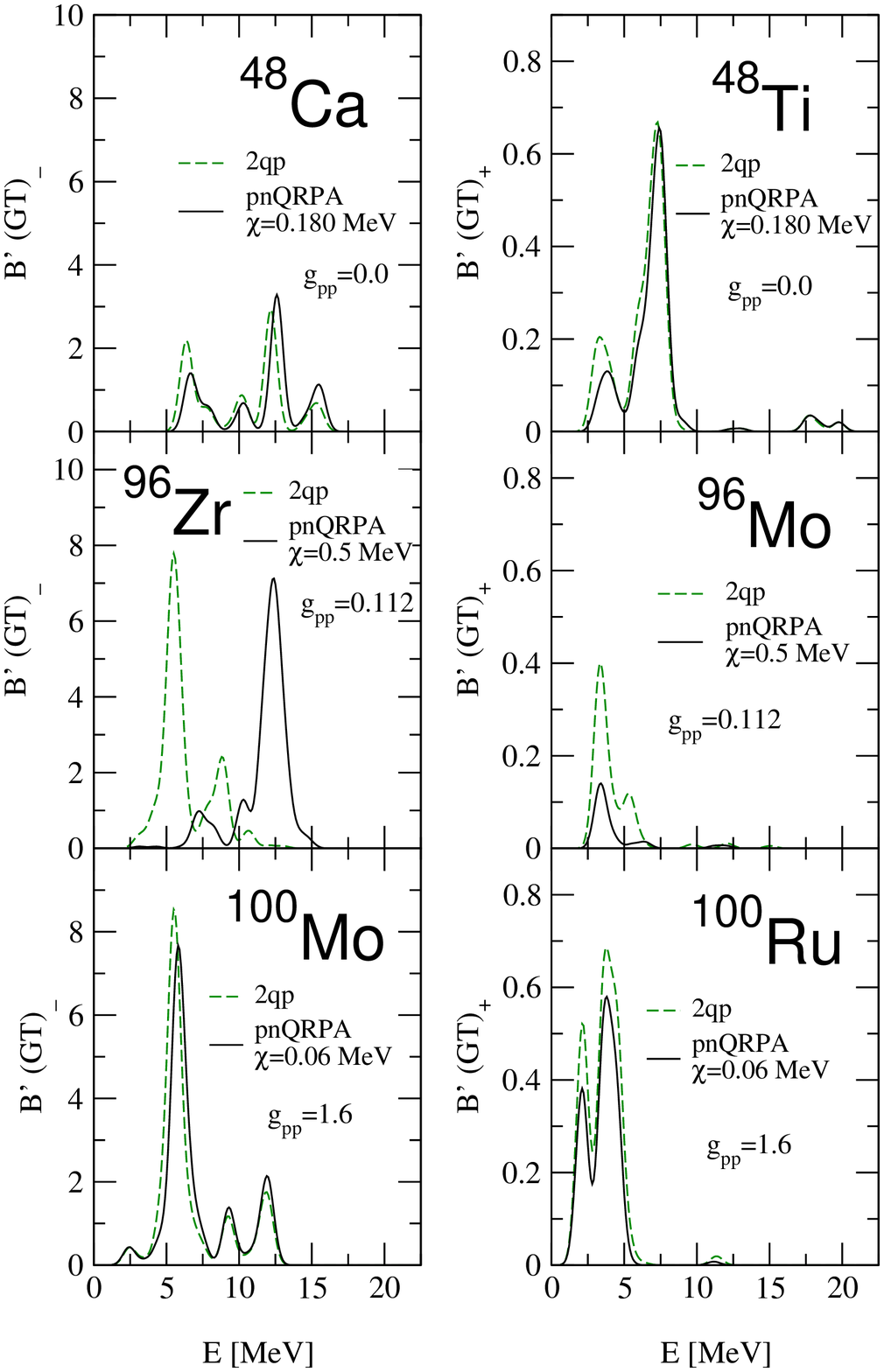,width=8cm,bbllx=5cm,%
bblly=10cm,bburx=18cm,bbury=26cm,angle=0}} 
\vskip8cm
\caption{(Color on line) Single $\beta^-$ strength, for $^{48}$Ca (upper-left panel),
$^{96}$Zr (middle-left panel), $^{100}$Mo (bottom-left panel),and single $\beta^+$ strength
for $^{48}$Ti (upper-right panel),$^{96}$Mo (middle-right panel),
$^{100}$Ru (bottom-right panel), folded with a Gaussian function having the
width of 1 MeV,
are plotted as a function of the energy within the BCS and pnQRPA approximation.
The pnQRPA calculations correspond to the values of $\chi$ and $g_{pp}$ listed
inside the graphs.}
\label{Fig. 1}
\end{figure}
\clearpage

\begin{figure}[h]
\centerline{\psfig{figure=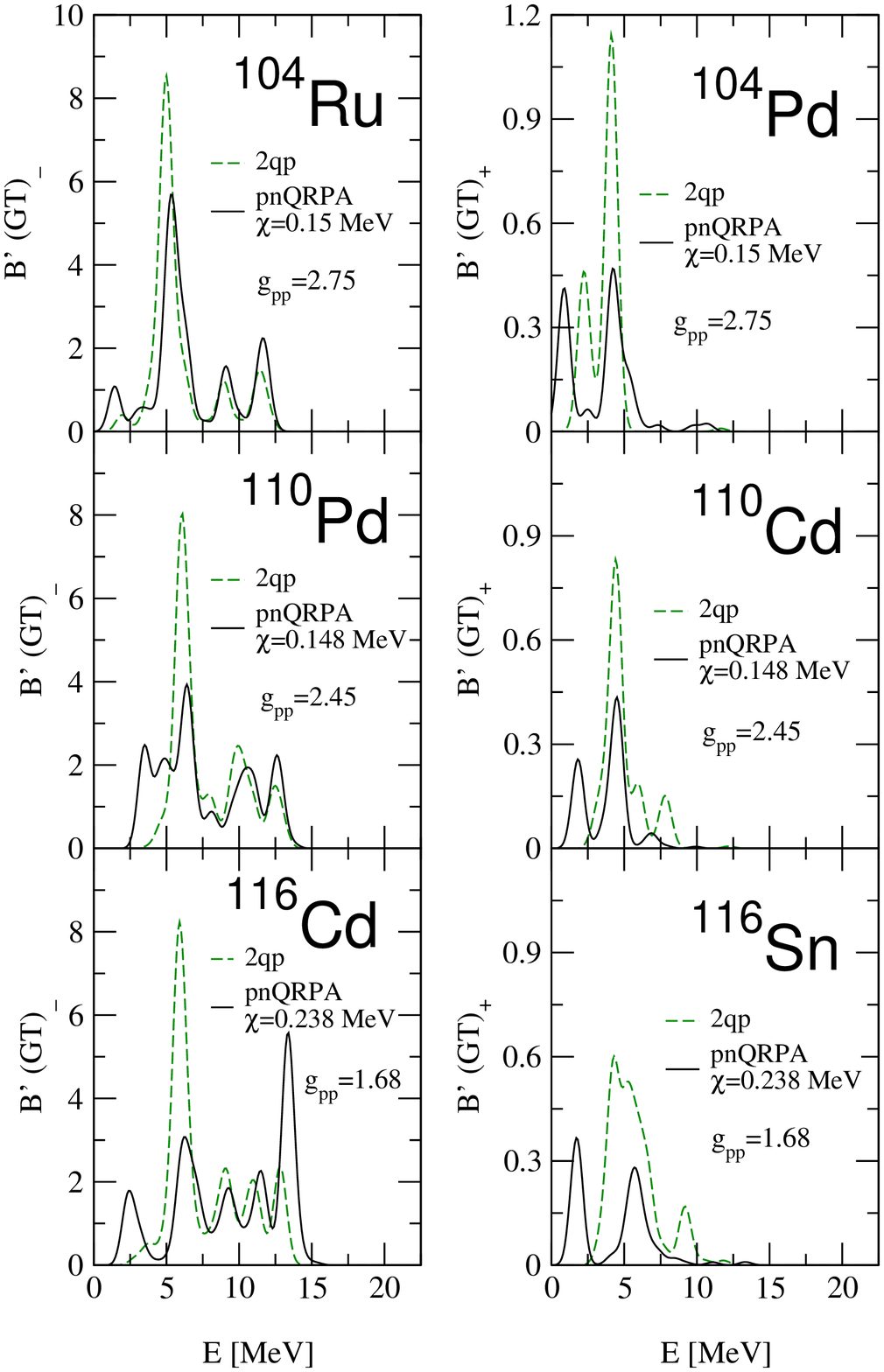,width=8cm,bbllx=5cm,%
bblly=10cm,bburx=18cm,bbury=26cm,angle=0}} 
\vskip8cm
\caption{(Color on line) Single $\beta^-$ strength, for $^{104}$Ru (upper-left panel),
$^{110}$Pd (middle-left panel), $^{116}$Cd (bottom-left panel),and single $\beta^+$ strength
for $^{104}$Pd (upper-right panel),$^{110}$Cd (middle-right panel),
$^{116}$Sn (bottom-right panel), folded with a Gaussian function having the
width of 1 MeV,
are plotted as a function of the energy within the BCS and pnQRPA approximation.
The pnQRPA calculations correspond to the values of $\chi$ and $g_{pp}$ listed
inside the graphs .}
\label{Fig. 2}
\end{figure}
\clearpage

\begin{figure}[h]
\centerline{\psfig{figure=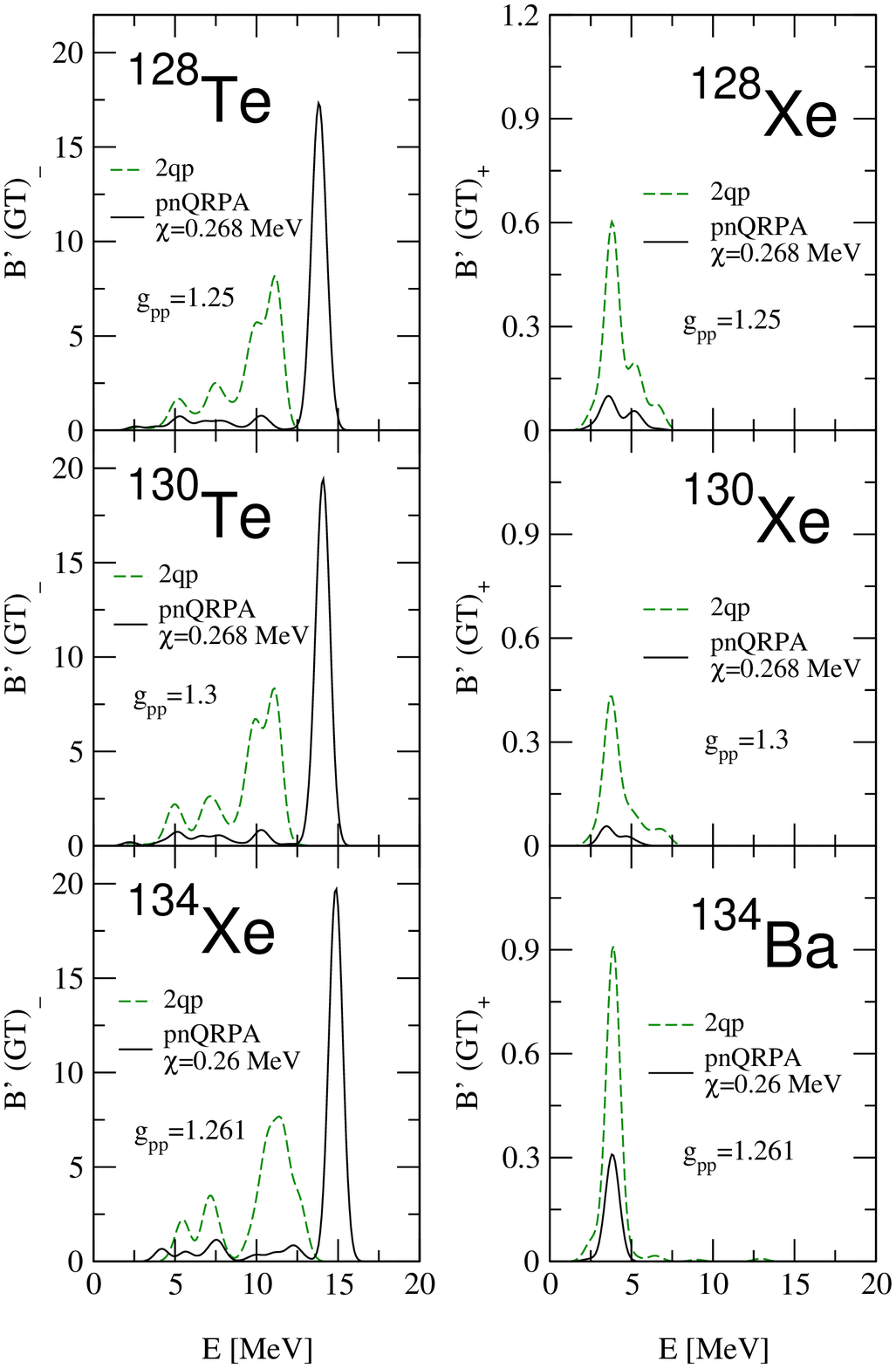,width=10cm,bbllx=5cm,%
bblly=10cm,bburx=18cm,bbury=26cm,angle=0}}
\vskip6.5cm
\caption{(Color on line) Single $\beta^-$ strength, for $^{128}$Te (upper-left panel),
$^{130}$Te (middle-left panel), $^{134}$Xe (bottom-left panel),and single $\beta^+$ strength
for $^{128}$Xe (upper-right panel),$^{130}$Xe (middle-right panel),
$^{134}$Ba (bottom-right panel), folded with a Gaussian function having the
width of 1 MeV,
are plotted as a function of the energy within the BCS and pnQRPA approximation.
The pnQRPA calculations correspond to the values of $\chi$ and $g_{pp}$ listed
inside the graphs .}
\label{Fig. 3}
\end{figure}
\clearpage

\begin{figure}[h]
\centerline{\psfig{figure=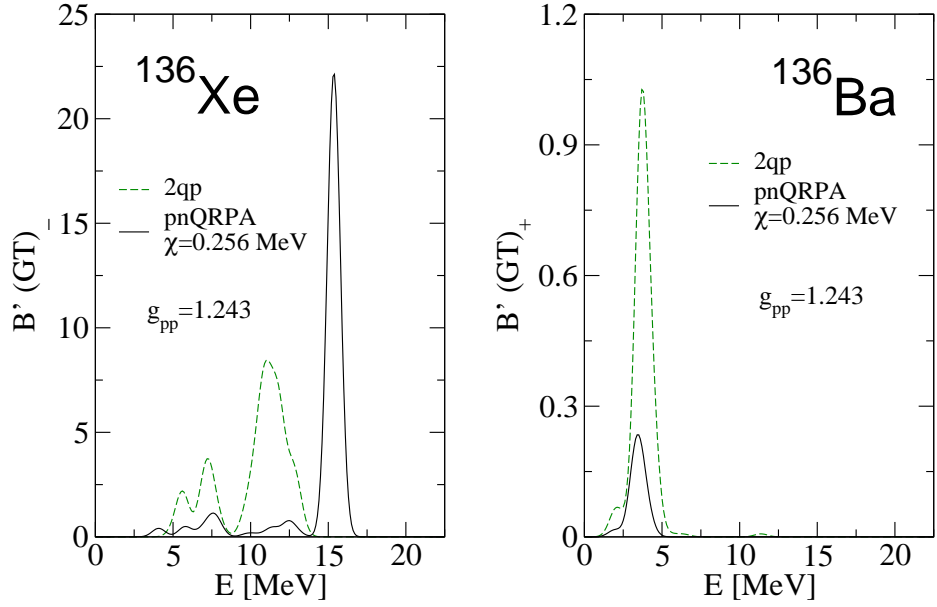,width=10cm,bbllx=3cm,%
bblly=5cm,bburx=18cm,bbury=26cm,angle=-90}}
\vskip8cm
\caption{(Color on line) Single $\beta^-$ strength, for $^{136}$Xe (left panel),
and single $\beta^+$ strength
for $^{136}$Ba (right panel), folded with a Gaussian function having the
width of 1 MeV,
are plotted as a function of the energy within the BCS and pnQRPA approximation.
The pnQRPA calculations correspond to the values of $\chi$ and $g_{pp}$ listed
inside the graphs.}
\label{Fig. 4}
\end{figure}
\clearpage

\begin{figure}[h]
\centerline{\psfig{figure=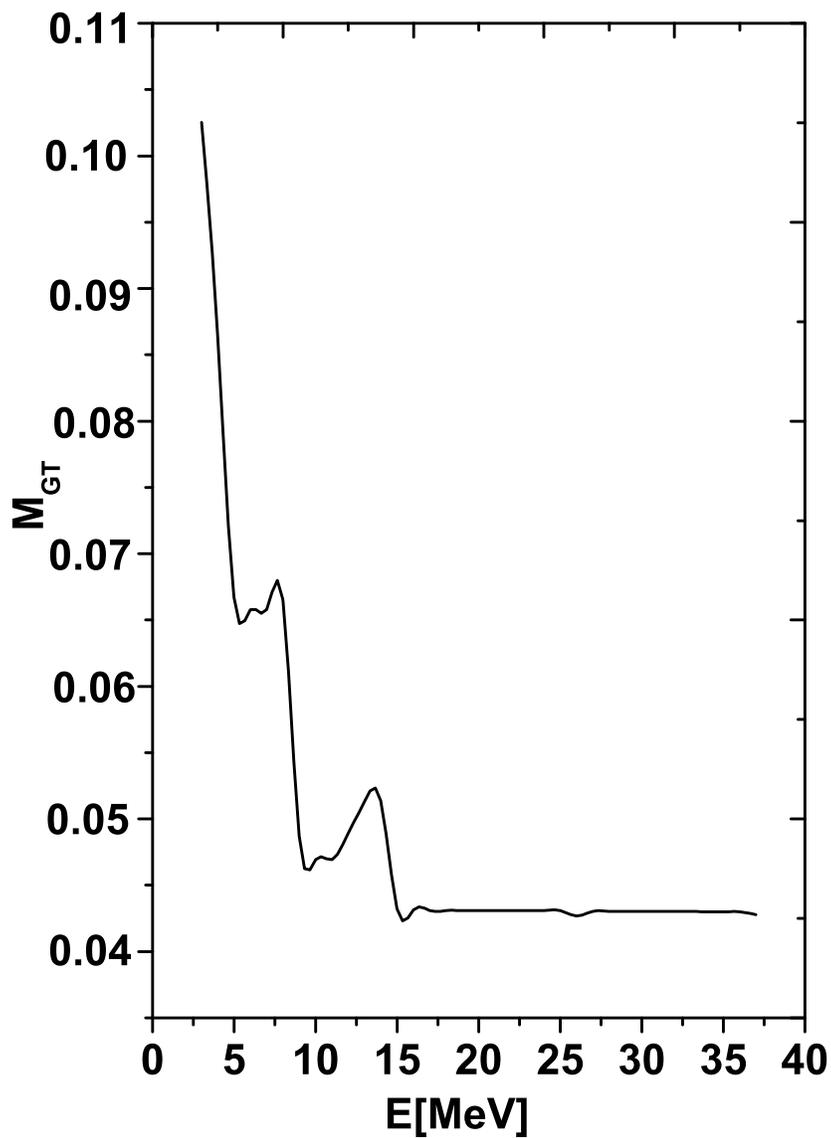,width=10cm,bbllx=3cm,%
bblly=5cm,bburx=18cm,bbury=26cm,angle=0}}
\vskip5cm
\caption{Double beta transition amplitude $M_{GT}$, given by Eq.(2.21), is
represented as function of energy E for $^{48}$Ca. The summation over k, in EQ. (2.21), is restricted
by $E_k\leq E$ where $E_k$ is defined by Eq. (2.22).
}
\label{Fig. 5}
\end{figure}
\clearpage
\end{document}